\documentclass{aa}
\usepackage[varg]{txfonts}
\usepackage{siunitx}
\usepackage{graphicx}
\usepackage{hyperref}

\bibpunct{(}{)}{;}{a}{}{,}

\hypersetup{ 
    colorlinks,
    linkcolor=blue,
    citecolor=blue
}

\newcommand{\intensity}{\unit{nW.m^{-2}.Hz^{-1}.sr^{-1}}}
\newcommand{\velocity}{\unit{km.s^{-1}}}
\newcommand{\colmass}{\unit{\kilogram.\metre^{-2}}}

\newcommand{\caII}{\ion{Ca}{ii}~}
\newcommand{\mgII}{\ion{Mg}{ii}~}

\begin{document}

\title{Spectral signatures of bright grains determine chromospheric heating}

\author{Elias R. Udn{\ae}s
        \inst{1,2}
        \and Tiago M. D. Pereira\inst{1,2}
        }
\institute{Rosseland Centre for Solar Physics, University of Oslo, P.O. Box 1029 Blindern, NO--0315 Oslo, Norway
\and
Institute of Theoretical Astrophysics, University of Oslo, P.O. Box 1029 Blindern, NO--0315 Oslo, Norway}
\date{}

\abstract
{Chromospheric heating is an important ingredient in the energy budget of the solar atmosphere. 
Quantifying chromospheric heating from observations is a challenging task because spectral observations suffer from non-LTE effects, and heating events may take place at spatial, temporal, and spectral resolutions difficult to achieve by modern telescopes.
By using 3D radiative magnetohydrodynamic simulations of the solar atmosphere combined with non-LTE spectral synthesis of several hundred snapshots, we estimated chromospheric heating from synthetic spectra and studied the spectral and temporal signatures of heating events. We performed k-means clustering on the \mgII h, \caII H, and \caII 8542~\AA\ lines to identify representative profiles associated with elevated chromospheric heating and studied their atmospheric stratification. We find that locations with the strongest chromospheric heating show spectral signatures with strong emission, typically the so-called chromospheric bright grains. Profiles with strong emission in the blue wing of the lines (blue grains) are created by upward-propagating shock waves and have an order of magnitude higher heating in the chromosphere than the ambient heating. Profiles with strong emission in the red wing (red grains) also display heating that is an order of magnitude stronger than the baseline, but these spectra do not show a characteristic atmospheric stratification. Spectra classified as blue grains have a consistent temporal evolution, which is an oscillating sawtooth pattern in the line core and emission in the blue wing. However, spectra classified as red grains did not show a consistent temporal signature: Red wing emission from the simulations can appear spontaneously or be associated with an oscillation. While red and blue grain profiles account for around 3\% of our synthetic spectra, they account for more than 12\% of the total chromospheric heating in these simulations. Around half of the chromospheric heating in the simulations came from locations where \caII H was in emission. By comparing two quiet Sun simulations, we find that the prevalence of bright grains is influenced by the magnetic field configuration, with a unipolar configuration showing fewer bright grains and consequently a lower share of heating from such events.}


\maketitle

\section{Introduction}

The chromosphere is a region of the solar atmosphere that is intrinsically dynamic. It is characterised by short-lived finely structured features such as fibrils, jets, and spicules. The complicated physics governing this region makes the chromosphere challenging to both model and interpret from observations. Processes in the chromosphere are driven by photospheric motions, and the chromosphere acts as a funnel of energy from the interior of the Sun to the corona. Therefore, determining energy transfer and dissipation through the chromosphere is necessary to understand how the hot corona is maintained.

In the chromospheric internetwork, steepening of acoustic waves creates rapid heating seen as bright grains in spectral lines such as \caII K \citep{Rutten:1991aa}. \citet[][]{Carlsson:1992vg} performed 1D numerical simulations of the acoustic wave propagation with non-LTE radiative transfer of the \caII K line, finding that waves with periods of \qty{180}{s} reproduce the bright grains seen in observations. Nonetheless, the heating contribution from these acoustic shock waves is still uncertain \citep{da-Silva-Santos:2024aa}, and accurate heating estimates are limited by several observational challenges such as spatial resolution \citep[noted in][]{Wedemeyer-Bohm:2007aa}, cadence, and non-LTE effects \citep{Carlsson:1992vg}. 

Improvements in solar instruments over the past decades have furthered our knowledge of the solar atmosphere, and brought new attention to the solar chromosphere \citep[see introduction in][]{Carlsson:2019aa}. Better spatially and spectrally resolved solar observations along with UV spectra of the chromosphere \citep{De-Pontieu:2014aa} have provided us with new opportunities to understand the dynamics and energy transfer in the solar atmosphere. Pushing the limits of solar observations requires a theoretical counterpart in order to understand and explain the energy transfer through the solar atmosphere. Often this theoretical framework is built upon 3D radiative magnetohydrodynamic (rMHD) simulations (e.g. MURaM, \citealt{Vogler:2005aa}; Bifrost, \citealt{Gudiksen:2005aa}; and CO5BOLD, \citealt{Freytag:2012aa}). Three-dimensional rMHD simulations have achieved a high amount of realism in the photosphere \citep{Pereira:2013ab}, and they can also mimic chromospheric dynamics and processes by including recipes for approximating non-LTE radiative losses \citep{Carlsson:2012aa} in codes such as Bifrost, or the more recent chromospheric extension to MURaM \citep{Przybylski:2022aa}. 

Using spectral observations to understand energy transfer through the chromosphere is complicated by the non-locality of the radiation formed in this layer. Therefore, studies of chromospheric heating often rely on simplified models such as the hydrostatic atmosphere, for example the famous VALC model \citep{Vernazza:1981aa} whose radiative losses are often taken as a flux to be balanced by heating processes \citep[e.g. the works of][]{Fossum:2005aa,Fossum:2006aa,Bello-Gonzalez:2009aa,Abbasvand:2021aa}, or they rely on non-LTE inversions in a 1.5D fashion \citep[e.g.][]{da-Silva-Santos:2024aa}. Another strategy is to use rMHD simulations to estimate heating of the chromosphere \citep{Molnar:2023aa,da-Silva-Santos:2024aa}. Forward modelling of spectral lines from simulations is computationally cheaper than non-LTE inversions, and it does not suffer from the degeneracy of the source function and linear extinction in the formal solution of radiative transfer that inversions do. By using spectra synthesised from 3D rMHD simulations, we can unambiguously link spectral signatures to atmospheric processes.

For this paper, we synthesised chromospheric spectral lines from 3D rMHD Bifrost simulations, obtaining a comprehensive dataset spanning hundreds of simulation snapshots. The spectra were used to identify signatures of chromospheric heating processes in quiet Sun conditions, and we estimate the total heating contribution coming from locations where we observe these signatures. We also identify the atmospheric properties and heating mechanisms behind these spectra, and study their temporal evolution.

\section{Methods}

\begin{figure*}
    \includegraphics[width=17cm]{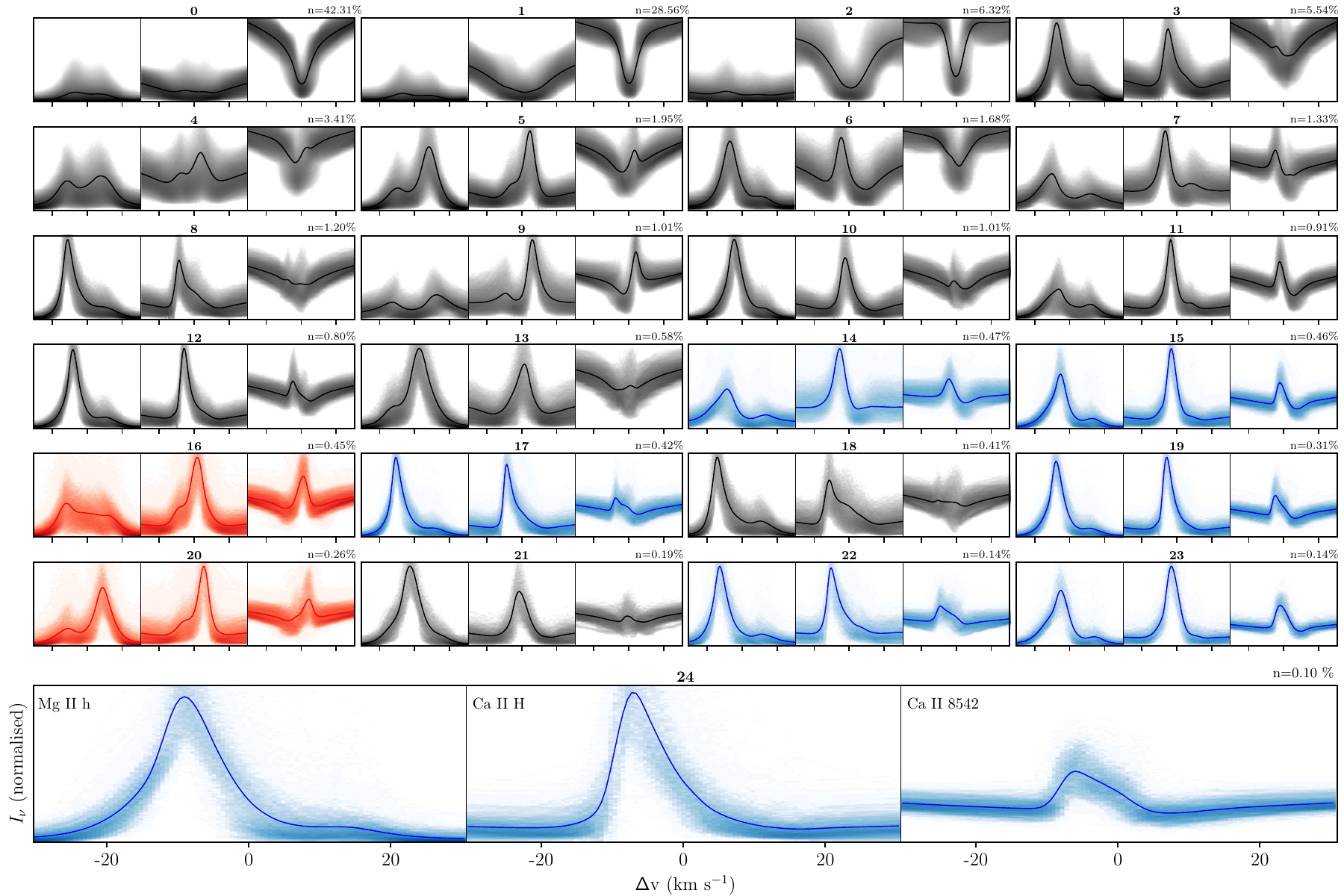}
    \caption{Simultaneous k-means clustering for the chromospheric \mgII h, \caII H, and \caII 8542~\AA~spectral lines. The clusters, labelled in bold, are ranked from most common in the top to least common in the bottom. The fraction of profiles in each cluster is labelled in the top right of each panel. The spectra are interpolated to a constant velocity separation for Doppler velocity between between $\pm$\qty{30}{\velocity}. Clusters of interest are divided into two classes: blue grains, plotted with blue; and red grains, plotted with red. In the figure, all clusters are normalised separately, but the scaling between the three lines in each cluster is preserved.}
    \label{fig:clustering}
\end{figure*}

\subsection{Simulations and synthetic spectra} \label{section:data}
To study the formation of chromospheric lines, we employed two 3D simulations produced with the Bifrost code. The simulations in our work cover the convection zone, photosphere, chromosphere, and corona, and we used recipes for non-LTE radiative losses in the chromosphere from \citet{Carlsson:2012aa}. Both simulations were magnetically quiet, with different magnetic field configurations mimicking quiet Sun conditions. One simulation, \texttt{qs024031}, was a typical quiet Sun atmosphere, and it is hereafter referred to as simulation A. The other simulation, \texttt{ch024031\_by200bz005}, has a unipolar photospheric magnetic field and vertical coronal magnetic field similar to that of a coronal hole, and it is referred to as simulation B. The simulation boxes were of identical size, with $768^3$ grid cells, a horizontal grid spacing of \qty{32.25}{\kilo\metre}, and a mean vertical grid spacing of \qty{21.9}{\kilo\metre}. At $z$=0, simulation B has a mean unsigned magnetic field of \qty{5.8}{mT}, and simulation A has \qty{4.1}{mT}. Simulation B has been used in several previous studies \citep[][]{Wedemeyer:2022aa,Silva:2022aa,Finley:2022aa,Brchnelova:2025aa,Udnaes:2025aa}. Simulation A is, as far as we know, has only been used in \citet[][]{Udnaes:2025aa}. 

While both simulations are magnetically quiet and can be thought of as quiet Sun simulations, their different magnetic topologies introduce changes in the atmosphere. Simulation B has a hotter corona and a transition region that occurs lower than for simulation A. The average chromospheric heating is also a few kilowatts per square metre higher in simulation B. These simulations do not have very different chromospheres, but the inclusion of two simulations gives us a broader range of temperatures and densities. Using two different simulations was a choice we made to get a better representation of quiet Sun profiles and explore heating signatures from different magnetic configurations.

The spectra used in this work were synthesised with the RH1.5D code \citep{Uitenbroek:2001wf,Pereira:2015wv} from a \qtyproduct{4 x 4}{\mega\metre} patch from both simulations. The spectra were calculated with a 10 s cadence for approximately 30 minutes from each simulation: from \qty{133}{minutes} until \qty{163}{minutes} for simulation A and from \qty{148}{minutes} until \qty{181}{minutes} for simulation B. We performed spectral synthesis for every second grid point in the horizontal directions to reduce the computation time, and because it has a negligible effect on the spatial resolution \citep[in Bifrost, the grid size typically oversamples the real resolution; see discussion in][]{Moe:2022aa}. The spatial sampling of the spectra used in this paper is then \qtyproduct{62.5 x 62.5}{\km}.

We focused on five chromospheric lines: \mgII h\&k, \caII H\&K, and \caII 8542~\AA. The lines were synthesised in non-LTE simultaneously using a six-level calcium model atom and a five-level magnesium model atom. The \mgII h\&k and \caII H\&K lines were modelled with partial redistribution (PRD), while the infrared \caII~8542~\AA~was modelled under the assumption of complete redistribution.

The near-UV lines \mgII h\&k are regularly observed by the Interface Region Imaging Spectrograph \citep[IRIS;][]{De-Pontieu:2014aa}, and their formation properties are described in a series of IRIS papers \citep{Leenaarts:2013aa,Leenaarts:2013ab,Pereira:2013aa}. The lines provide diagnostics of the upper chromosphere, with the full lines tracing heights from the upper photosphere. Upward-propagating shocks are seen in spectrograms of \mgII h \citep{De-Pontieu:2015aa}, making this line suitable for our analysis. Synthetic \mgII h\&k line profiles from Bifrost simulations are narrower than observations \citep{Pereira:2013aa}, which \citet{Hansteen:2023aa} suggest could originate from a lack of mass loading into the chromosphere. \citet{Ondratschek:2024aa} used a MURaM-ChE simulation and report synthetic \mgII h\&k profiles that are broader than those from Bifrost but still not as broad as observed by IRIS. The authors suggest that insufficient dynamics in the simulations may explain the difference. The specific reason for the narrower synthetic \mgII\ profiles is of no concern for this work. The mechanisms for line formation and energy deposition are still general, even if our simulations display fewer broad \mgII h\&k profiles when compared with observations. 

The formation of the \caII H\&K lines is similar to that of the \mgII h\&k lines, but Ca is less abundant in the solar atmosphere \citep{Asplund:2009aa} and thus formed at lower heights. \caII H\&K are in the visible part of the spectrum and therefore observable with ground-based instruments, for example the CHROMIS instrument in the Swedish 1-meter solar telescope \citep[SST;][]{Scharmer:2003aa}. These lines provide good diagnostics for chromospheric heating and show bright grains associated with acoustic shocks \citep{Carlsson:1992vg}. The \caII H\&K lines are formed at heights where 3D effects could be important for the line cores. \cite{Bjorgen:2018aa} show that PRD effects are more important than 3D effects in the line wings, and 1.5D modelling of the lines under PRD is a feasible option for the wings. \citet{Wedemeyer-Bohm:2011aa} also show that statistical equilibrium is a good approximation for \caII lines.

The \caII 8542~\AA~line is one of calcium's infrared triplet lines, formed lower than \caII H\&K. This line is a popular choice for inferring the chromospheric magnetic field \citep[e.g.][]{de-la-Cruz-Rodriguez:2012aa, Quintero-Noda:2016aa, Kuridze:2017aa}, but it also shows the bright grain structure associated with acoustic shock waves \citep{Leenaarts:2009aa, Martinez-Gonzalez:2023aa}.

\subsection{Clustering} \label{section:clustering}

We employed clustering to find correlations between synthetic spectra and chromospheric heating in the simulations. This was necessary given the sheer number of spectra and to get a statistical view of chromospheric heating. We used k-means clustering to extract the shapes of the spectra. This algorithm was chosen for its speed and simplicity, making it possible to perform several experiments for different initialisation parameters of the algorithm.

K-means clustering is an unsupervised machine learning technique that is routinely used to analyse spectral data. K-means has been used to analyse both solar observations \citep[e.g.][]{Joshi:2020aa,Barczynski:2021aa,Nobrega-Siverio:2021aa,Woods:2021aa,Kleint:2022aa} and synthetic spectra from simulations \citep{Moe:2023aa,Moe:2024aa,Mathur:2025aa}. The algorithm groups data into $k$ clusters based on similarity, where the similarity is defined by the $L_2$ norm. Formally, k-means selects clusters by trying to minimise the within-cluster sum of squares,
\begin{equation}
    \sum_i || I_{\lambda,i} - I_{\lambda,c} || ^2\,,
\end{equation}
where $I_{\lambda,c}$ is the centroid of the cluster. This centroid is computed as the average spectral profile of a cluster, and is often referred to as the representative profile in spectral line clustering. 

In this work, we clustered the \mgII h, \caII H, and \caII~8542~\AA~lines simultaneously, by concatenating the lines. Stacking spectral information from three lines was found to better constrain the underlying atmosphere and resulted in better estimates of chromospheric heating and stratifications. Clustering different lines simultaneously is not a new idea, and has been done before in observational studies \citep[e.g. clustering Stokes components of \ion{Fe}{i} lines][]{Khomenko:2003aa,Pietarila:2007aa,Viticchie:2011aa}.

Before clustering the spectra, we performed three important pre-processing steps. We interpolated the spectra into a wavelength grid with constant spacing, normalised the spectra separately for each of the three lines, and performed feature selection in order to favour clusters with emission signatures. 

Firstly, we interpolated the spectra into a wavelength grid with constant spacing, to give equal weight to each wavelength bin in the clustering. We restricted the wavelength ranges to $\pm\qty{30}{\velocity}$ around the line core for each line, which discards the continuum and concentrates only on the chromospheric part of the lines. 

Secondly, we performed Z-normalisation for each line separately, to give all three lines the same weight in the distance metric in k-means. Otherwise, the clusters would have poor sensitivity to the \mgII h line, which usually is much fainter than the \caII lines. Z-normalisation preserved spectral line information with respect to the spectral line mean and standard deviation. 

Lastly, we filtered out the most common profiles to make the clustering more sensitive to the spectral signatures of heating. In the quiet Sun chromosphere, most \caII line profiles are in absorption. For our purposes, absorption profiles are less relevant since the strongest heating events usually show emission in \caII\ lines. Therefore, we discarded most of the spectra where the \caII lines were in absorption before clustering the spectra. This was done through an initial clustering with only eight clusters. We selected the clusters where the centroids of the \caII lines were in absorption, and removed 90\% of the profiles from these these clusters. This left us with 20\% of the total spectra, where ca. 60\% of the profiles had central emission in the \caII lines. Reducing the amount of absorption profiles allowed us to capture more variation in clusters of emission profiles while using fewer clusters.
\begin{table}
    \centering
    \caption{Peak intensities of representative profiles.}
    \begin{tabular}{c r r r r} 
    \hline\hline
    Cluster & \mgII $\mathrm{k_{2V}}$ & \mgII $\mathrm{k_{2R}}$ &  \caII H & \caII 8542~\AA\\
            no. &    (int. units)         &     (int. units)        &    (int. units)     &       (int. units)      \\[0.5ex] 
    \hline
    0 & 0.91 & 0.74 & $1.09$ &  --- \\
    1 & 0.92 & 0.70 & --- &  --- \\
    2 & 1.29 & 0.89 & --- &  --- \\
    3 & 7.74 & 1.62 & $7.15$ &  $5.44$ \\
    4 & 3.29 & 3.84 & $6.43$ &  $6.99$ \\
    5 & 2.63 & 7.42 & $9.30$ &  $7.03$ \\
    6 & 8.17 & 1.68 & $8.59$ &  --- \\
    7 & 5.95 & 2.69 & $12.90$ &  $9.77$ \\
    8 & 11.98 & --- & $8.90$ &  $5.97$ \\
    9 & 2.97 & 4.24 & $13.80$ &  $11.69$ \\
    10 & 12.68 & --- & $9.82$ &  $6.13$ \\
    11 & 5.73 & 2.84 & $15.05$ &  $10.92$ \\
    12 & 14.77 & 2.08 & $14.98$ &  $8.86$ \\
    13 & 11.48 & --- & $9.20$ &  $6.14$ \\
    14 & 10.68 & --- & $21.74$ &  $13.43$ \\
    15 & 14.47 & 2.58 & $21.29$ &  $12.19$ \\
    16 & 6.81 & 4.37 & $16.05$ &  $12.23$ \\
    17 & 20.51 & 2.36 & $18.74$ &  $10.08$ \\
    18 & 15.68 & 3.15 & $11.21$ &  $7.22$ \\
    19 & 22.24 & 2.59 & $23.35$ &  $12.24$ \\
    20 & 4.24 & 14.19 & $19.38$ &  $11.40$ \\
    21 & 20.67 & --- & $14.13$ &  $7.76$ \\
    22 & 25.79 & 3.84 & $25.27$ &  $12.91$ \\
    23 & 21.61 & 3.58 & $31.07$ &  $15.81$ \\
    24 & 31.24 & 3.33 & $32.09$ &  $15.23$ \\

    \hline
    \end{tabular}

    \label{tab:RP_int}

    \tablefoot{Intensity (int.) units are \intensity. Some representative profiles did not have a discernable peak.}
\end{table}

\begin{table}
    \centering
    \caption{Velocity shifts of representative profiles.}
    \begin{tabular}{c r r r r} 
    \hline\hline
    Cluster & \mgII $\mathrm{k_{2V}}$ & \mgII $\mathrm{k_{2R}}$ &  \caII H & \caII 8542~\AA\\
          no.  &    (\velocity)         &     (\velocity)        &    (\velocity)     &       (\velocity)      \\[0.5ex] 
    \hline
    0 & $-6.6$ & 6.6 & $-4.8$ &  --- \\
    1 & $-6.6$ & 7.8 & --- &  --- \\
    2 & $-9.0$ & 8.4 & --- &  --- \\
    3 & $-7.2$ & 9.6 & $-4.8$ &  $-4.2$ \\
    4 & $-11.4$ & 7.8 & $3.6$ &  $3.6$ \\
    5 & $-9.6$ & 7.8 & $4.8$ &  $3.6$ \\
    6 & $-7.2$ & 7.8 & $-4.8$ &  --- \\
    7 & $-10.2$ & 11.4 & $-6.6$ &  $-4.8$ \\
    8 & $-10.8$ & --- & $-8.4$ &  $-7.8$ \\
    9 & $-12.6$ & 12.6 & $6.0$ &  $4.2$ \\
    10 & $-4.8$ & --- & $-2.4$ &  $-1.2$ \\
    11 & $-6.0$ & 11.4 & $-3.6$ &  $-2.4$ \\
    12 & $-7.8$ & 10.2 & $-5.4$ &  $-4.8$ \\
    13 & $2.4$ & --- & $1.8$ &  $1.2$ \\
    14 & $-8.4$ & --- & $-5.4$ &  $-4.2$ \\
    15 & $-5.4$ & 12.6 & $-3.0$ &  $-2.4$ \\
    16 & $-11.4$ & 10.8 & $1.8$ &  $1.2$ \\
    17 & $-10.8$ & 9.6 & $-8.4$ &  $-7.2$ \\
    18 & $-13.8$ & 10.2 & $-11.4$ &  $-10.2$ \\
    19 & $-7.8$ & 10.8 & $-5.4$ &  $-4.8$ \\
    20 & $-10.8$ & 9.0 & $5.4$ &  $4.2$ \\
    21 & $-2.4$ & --- & $-1.2$ &  $-0.6$ \\
    22 & $-12.6$ & 10.8 & $-10.2$ &  $-9.0$ \\
    23 & $-4.8$ & 13.8 & $-3.0$ &  $-2.4$ \\
    24 & $-9.0$ & 12.0 & $-7.2$ &  $-6.0$ \\

    \hline
    \end{tabular}

    \label{tab:RP_vel}

\end{table}

\begin{figure}
    \resizebox{\hsize}{!}{\includegraphics{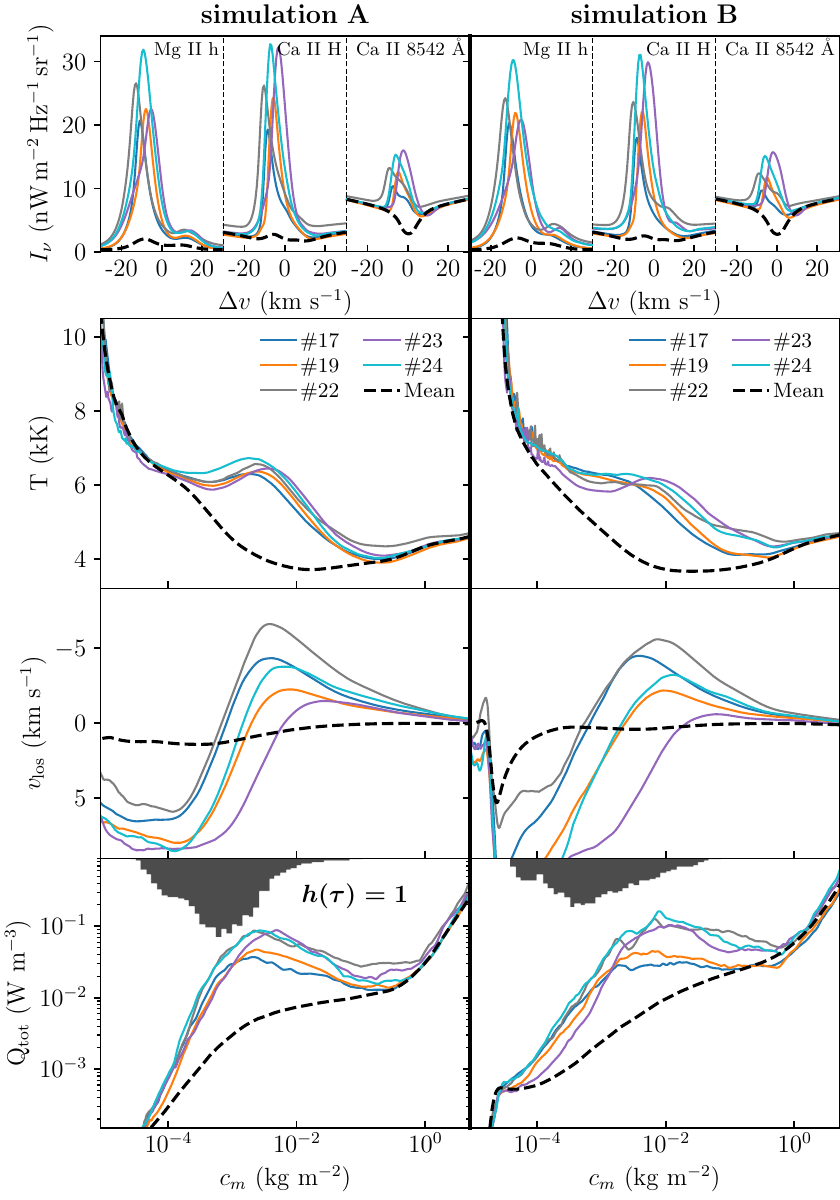}}
    \caption{Atmospheric profiles for the clusters with high emission in the blue wing of the representative profiles. The columns are split into the two different simulations (left: simulation A, right: simulation B). Top row: Average spectra of the clusters. Bottom three rows: Average atmospheric quantities in the clusters. From the top, they are the gas temperature, line-of-sight velocity, and total dissipative heating ($Q_\mathrm{viscous} + Q_\mathrm{joule}$). The dashed black lines show the averages for the entire simulations. The distribution in the bottom row is the column mass of the blue emission clusters where $\tau_\nu$ equals unity for the inner blue wing (\qty{-5.4}{\velocity}) of \caII H.}
    \label{fig:cluster_atmos_blue}
\end{figure}

\begin{figure}
    \resizebox{\hsize}{!}{\includegraphics{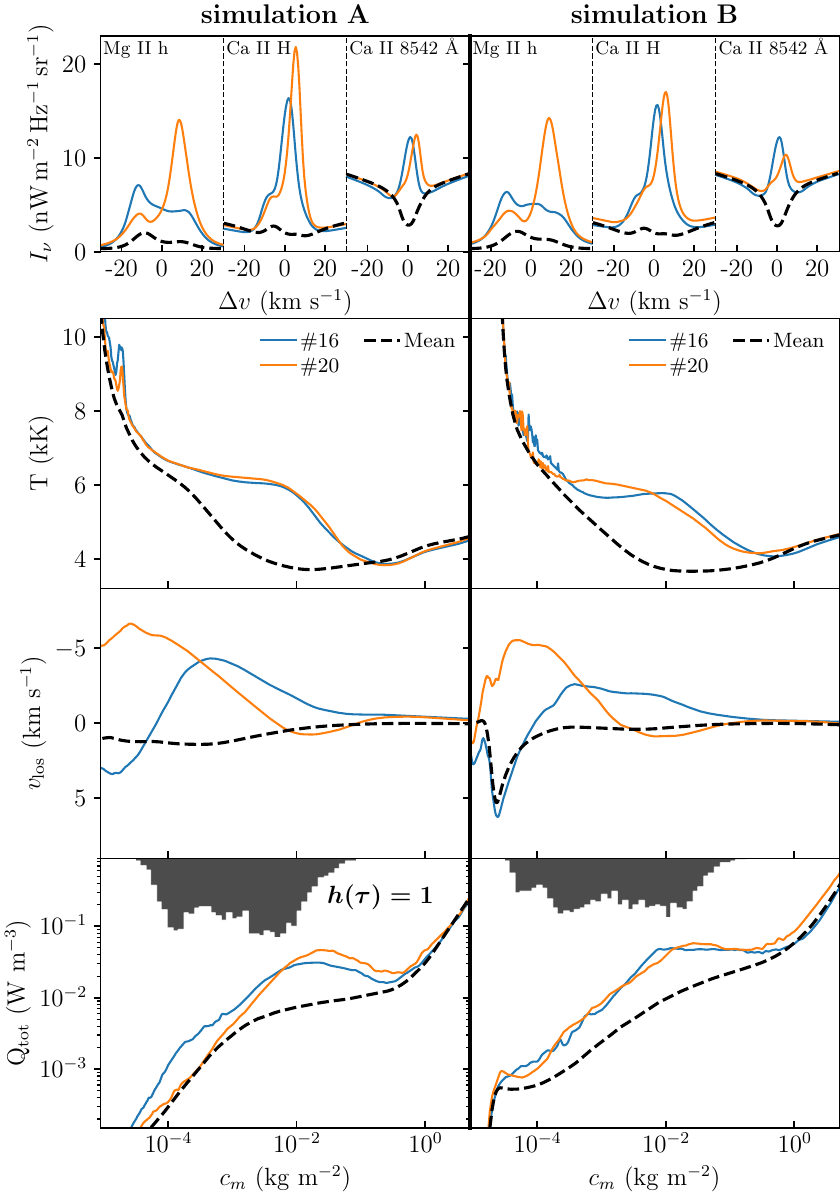}}
    \caption{Atmospheric profiles for the clusters with high emission in the red wing of the representative profiles. See caption of Fig.~\ref{fig:cluster_atmos_blue}. Here, the distribution in the bottom row is the column mass of the red emission clusters where $\tau_\nu$ equals unity for the inner red wing (\qty{5.4}{\velocity}) of \caII H.}
    \label{fig:cluster_atmos_red}
\end{figure}

\section{Results}

\subsection{Clustering}

We clustered the \mgII h, \caII H, and \caII~8542~\AA~lines from the two Bifrost simulations described in section \ref{section:data}. The number of clusters was chosen to minimise both the variation within each cluster and the number of clusters. A good rule of thumb is to use the inflection point of inertia as the number of clusters via the elbow method. In our analysis, there was no clear elbow in the inertia plot (see Fig~\ref{fig:inertia}), which is common for complex data such as synthetic spectra. We saw a good correspondence between cluster variation and representative profiles for $\gtrsim20$ clusters, and after reducing the absorption profiles from the dataset (described in section~\ref{section:clustering}), we clustered the data with 25 clusters. After the k-means algorithm converged, we obtained representative profiles of the reduced dataset that gave more weight to emission profiles. Using the representative profiles of the reduced data, we clustered the discarded absorption profiles in order to include the entire dataset of spectra in our analysis. 87.5\% of the discarded profiles were placed in clusters where both calcium lines were in pure absorption. Using different initialisations or a larger number of clusters had a negligible effect on our results, and we discuss this in more detail in Appendix~\ref{app:B}.

Figure~\ref{fig:clustering} shows the representative profiles along with the spectral distribution of each cluster. The clusters are sorted by size, meaning that cluster 0 is the most common and cluster 24 the least common. Cluster 0 has 42\% of the spectra, while cluster 24 has only 0.1\% of the spectra. The large amount of samples in cluster 0 and very small amount of samples in the last cluster, is both a consequence and a desired effect of the data filtration we performed before clustering the data. In Fig.~\ref{fig:clustering}, the most common clusters show typical absorption profiles in the \caII lines and weak emission in the \mgII h2V and h2R peaks. Many of the less common clusters show strong emission in the core of the calcium spectral lines, which is typically associated with chromospheric heating.

We find that spectra with the strongest emission are often associated with strong Doppler shifts and shock waves.
Following the naming of \citet{Beckers:1964aa}, we call the profiles with the strongest emission by `grains' (blue or red-shifted). Clusters 14, 15, 17, 19, 22, 23, and 24 therefore correspond to bright blue grains, where the \caII H peak is above \qty{15}{\intensity} and the \caII 8542~\AA~peak is stronger than \qty{10}{\intensity}. Clusters 16 and 20 correspond to bright red grains with the same peak intensity thresholds as for the bright blue grains. The peak intensity values are summarised in Table~\ref{tab:RP_int}, and the velocity shift of the emission peaks are summarised in Table~\ref{tab:RP_vel}. Most of the \caII K lines in strong emission display only one peak, and therefore we only record the strongest peak for this line.

\begin{figure}
    \resizebox{\hsize}{!}{\includegraphics{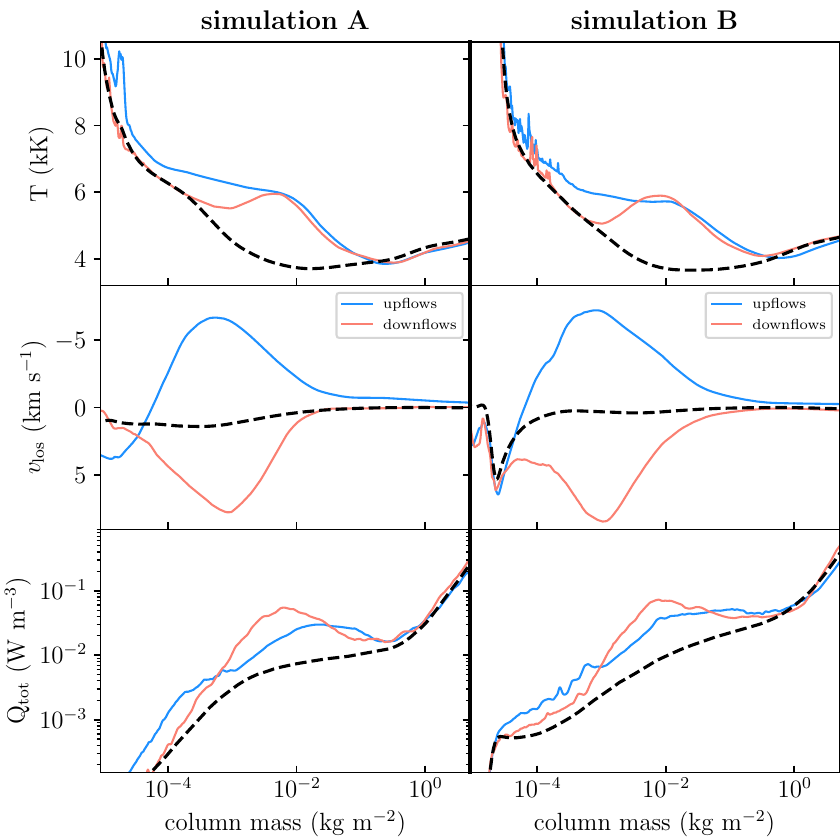}}
    \caption{Two typical atmospheres for cluster 16: one atmosphere with down-flow and one with up-flow in the chromosphere. The blue lines are average quantities for all pixels with a negative line-of-sight velocity at $c_m = 10^{-3}$ and the red lines are average quantities for all pixels with a positive line-of-sight velocity at the same column mass. We show the $\tau = 1$ distributions for the up-flowing (blue) and down-flowing (red) at a Doppler velocity of \qty{+5.4}{\velocity} for the \caII H line.}
    \label{fig:cluster16}
\end{figure}

By averaging the atmospheric quantities per columns mass for each cluster, we calculated the `typical' atmosphere for each cluster, and for each simulation. The clusters we were interested in were the clusters with strong emission, i.e. the blue and red grains. The atmospheric profiles for the representative profiles with brightest blue grains are presented in Fig.~\ref{fig:cluster_atmos_blue}. The top row of this figure shows the average spectra for each cluster. Because we calculated the average spectra and atmospheric quantities separately for the two simulations, the spectra in the first row of the figure are slightly different than the representative profiles shown in Fig.~\ref{fig:clustering}. The bottom three rows in Fig.~\ref{fig:cluster_atmos_blue} show the average atmospheric profiles for the clusters with high emission in the blue wing. While Fig.~\ref{fig:cluster_atmos_blue} shows some differences in atmospheres between the simulations, the trends are the same. Compared to the average, the temperature minima of bright blue grain atmospheres are hotter, occur at higher column masses, and are followed by temperature rises to more than \qty{6000}{K}, around \qty{2500}{K} hotter than the mean temperature at the same column mass in the chromosphere. The line-of-sight velocities show a reversal from up-flowing to down-flowing plasma. This is the shock front created by acoustic waves leaking from the convection zone. The shock front creates an excess of chromospheric heating, seen in the bottom row of the figure. In the same panel, the $\tau_\nu = 1$ distribution shows that most of the profiles form near the shock-front, where the chromospheric temperature peak is located.

Figure~\ref{fig:cluster_atmos_red} shows the typical atmospheres for bright red grains. These clusters have similar temperature profiles to the clusters with blue-shifted emission, but different line-of-sight velocities. Many atmospheric profiles in cluster 20 have a down-flow at a low height in the atmosphere, where the red emission peak is formed. This can be seen as a weak negative line-of-sight velocity in Fig.~\ref{fig:cluster_atmos_red}, but the average line-of-sight velocity is not corresponding to the velocity shift of the calcium emission peaks (seen in Table~\ref{tab:RP_vel}). The average line-of-sight velocity for cluster 16 does not show any down-flow around the formation height of the line. 

Cluster 16 shows broad representative profiles with strong emission peaks that are slightly red-shifted in the \caII lines. The \mgII h representative profile does however have  a stronger h2V than a h2R peak. Analysing the individual atmospheres of these profiles, we could not identify a common pattern in how the lines form. As seen in Fig.~\ref{fig:cluster_atmos_red}, the height of formation is spread over a large range, and the average line-of-sight velocity is mostly positive around the height of formation for \caII H. Upon further inspection of their atmospheric stratifications, we found that the average line-of-sight velocity is poorly constrained for both clusters 16 and 20, unlike for the bright blue grains. For cluster 16, the atmospheres fall into two main distributions: a typical atmosphere where there is a chromospheric up-flow of about \qty{7}{\velocity}, and a less typical atmosphere with a chromospheric down-flow of about \qty{8}{\velocity}. A break-down of the two different components is shown in Fig.~\ref{fig:cluster16}. For the up-flowing atmospheres, the red emission peak is formed lower in the atmosphere, at a column mass of around \qty{0.1}{\colmass}. Above the formation of the red peak, there is a strong up-flow which does not absorb the intensity in the red. The blue part of the line is formed much higher in the chromosphere, between \qty{1e-3}{\colmass} and \qty{1e-4}{\colmass}. For the down-flowing atmospheres, the formation of the red peak forms mainly in the down-flow. For simulation A, only 17\% of the profiles in cluster 16 have a down-flow in the chromosphere, while for simulation B this fraction is 31\%.

\begin{figure}
    \resizebox{\hsize}{!}{\includegraphics{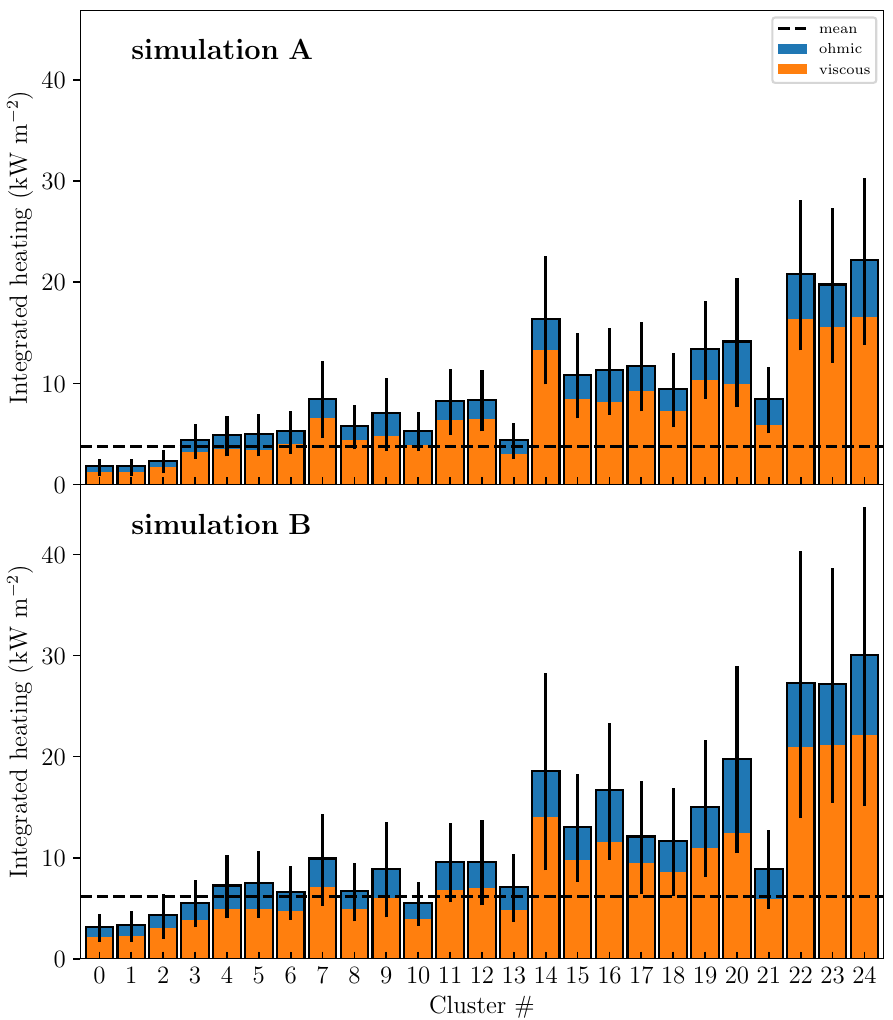}}
    \caption{Median of the integrated chromospheric viscous and Joule heating per cluster. Vertical black lines give the median absolute deviation, i.e. the spread of values inside each cluster. The dashed horizontal lines are the averages of the integrated chromospheric heating ($Q_{\nu} + Q_{\eta}$) for each simulation.}
    \label{fig:cluster_heating}
\end{figure}

\subsection{Heating signatures and temporal evolution}

From the clustering we established that the strongest spectral signatures of chromospheric heating are associated with bright grains. We then focused on two objectives: quantifying the proportion of total chromospheric heating associated with bright grains; and characterizing the temporal evolution of spectra arising from the heating events.

To quantify the energy dissipated in the chromosphere, we used the viscous and Joule heating terms from the simulations, $Q_{\nu} + Q_{\eta}$ (we refer to the discussion and appendix E in \citealt{Noraz:2025aa} for a discussion of the terms in the energy equation that are dissipative). The total chromospheric heating was calculated by integrating the sum of the volumetric viscous and Joule dissipation terms from the simulation column-by-column through the chromosphere. The result is an energy flux of total dissipative energy from viscous and Joule heating in the chromosphere. To avoid including photospheric heating, we limited the integration to a range in column mass. The lower limit was \qty{0.1}{\colmass}, which was between \qtyrange{600}{800}{\kilo\metre} in height for most columns. The upper limit was \qty{2e-5}{\colmass} for simulation B (average height of \qty{3.1}{\mega\metre}) and \qty{3e-6}{\colmass} for simulation A (average height of \qty{5.2}{\mega\metre}). We chose a higher column-mass limit for simulation B since its transition region appears at a higher column mass (and lower height) than for simulation A. In the simulations, strong heating events represented a large fraction of the total chromospheric heating, with 15\% of the columns responsible for around half of the total dissipative heating. Cumulative distributions of chromospheric heating are shown in Fig.~\ref{fig:cumulative_heating} of Appendix~\ref{app:A}.

Figure~\ref{fig:cluster_heating} shows the distribution of chromospheric heating per cluster. The figure shows a clear correlation between clusters with line core emission and chromospheric heating. Clusters 22, 23, and 24 have the strongest emission peaks for the three lines and are also the clusters where the chromosphere has the highest total heating.  Viscous heating is the highest for all clusters, which is expected in a quiet Sun region with weak magnetic fields. 

\begin{figure*}
    \sidecaption
    \includegraphics[width=12cm]{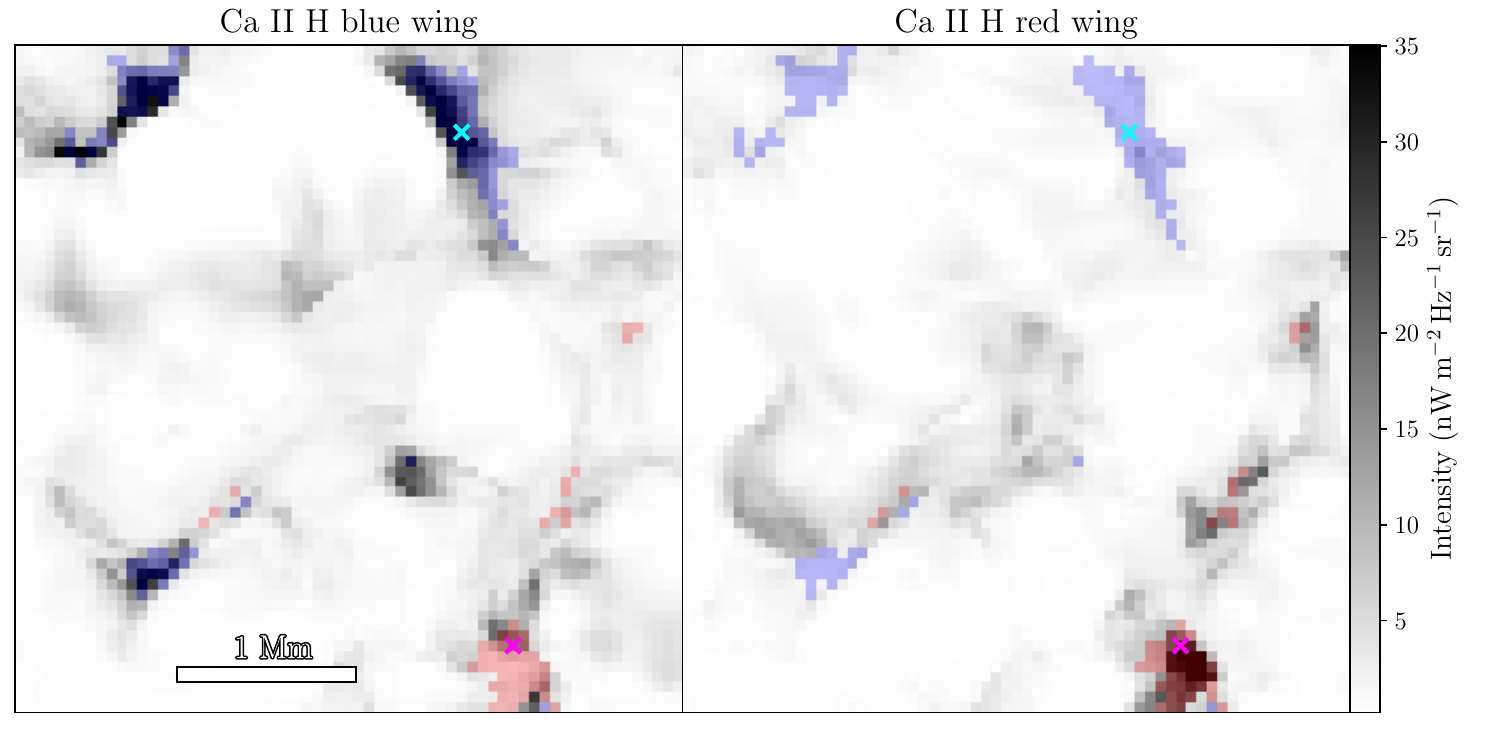}
    \caption{Images of the \caII H blue wing at \qty{-5}{\velocity} and red wing at \qty{+5}{\velocity} from simulation A. Overplotted on the images are locations where the pixels belong to a cluster with strong blue emission (cluster 14, 15, 17, 19, 22, 23, and 24) in blue or strong red emission (cluster 16 and 20) in red. The cyan and magenta crosses are placed on locations with especially strong blue or red emission, respectively. The pixel marked with cyan belongs to cluster 24, and the pixel marked with magenta belongs to cluster 20.  Spectrograms from these pixels can be seen in Fig.~\ref{fig:temporal_clustermap}.}
    \label{fig:cluster_map}
\end{figure*}

\begin{figure*}
    \includegraphics[width=17cm]{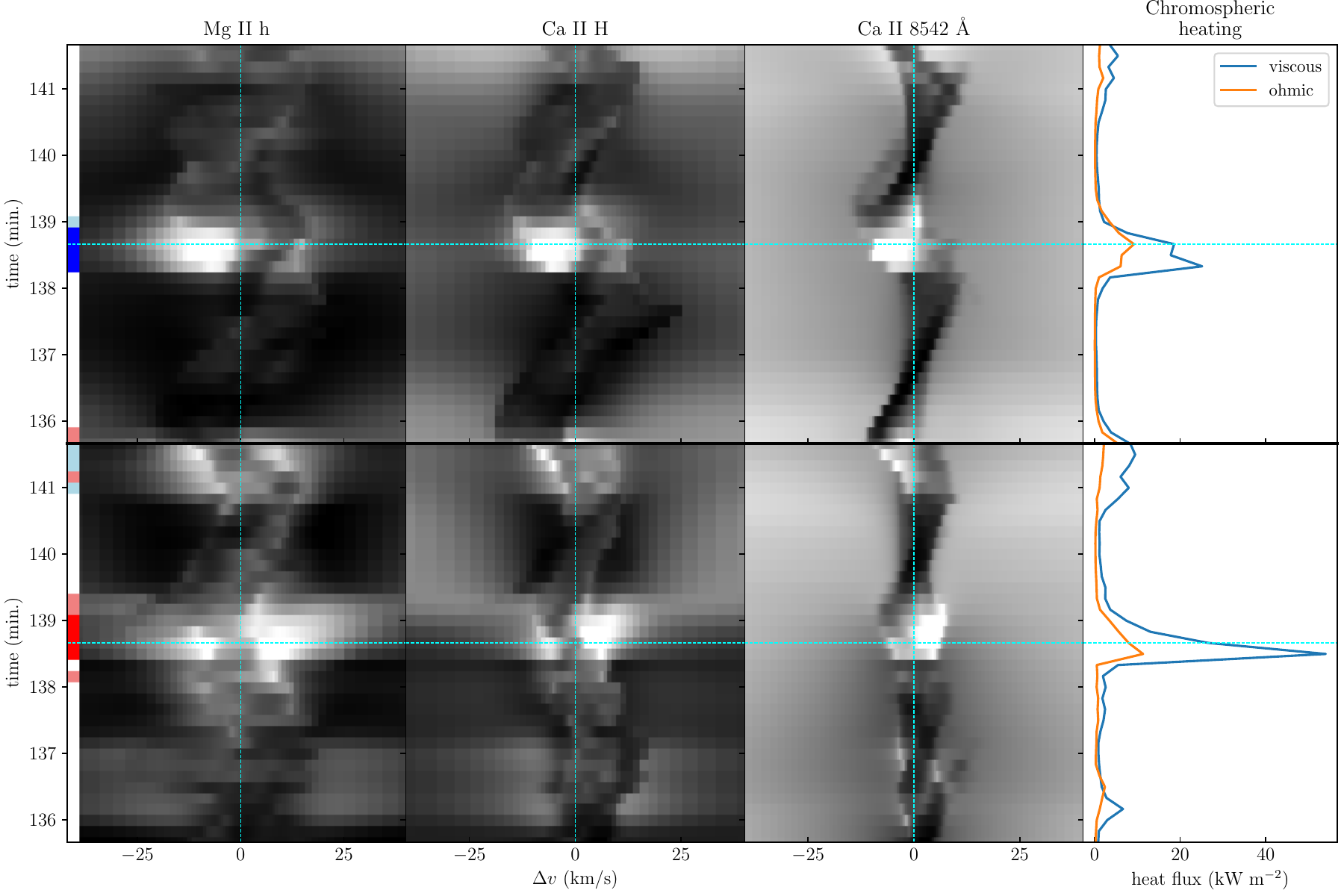}
    \caption{Spectrograms of two heating events from simulation A. Top row: bright blue grain from the pixel that is marked with a cyan cross in Fig.~\ref{fig:cluster_map}. Bottom row: bright red grain from the pixel that is marked with a magenta cross in Fig.~\ref{fig:cluster_map}. Both events display an excess of chromospheric heating, shown in the column to the right.  The spectrograms are scaled non-linearly with a $\gamma$ factor to enhance contrast. For \mgII h and \caII H, $\gamma = 1/4$; for \caII 8542~\AA, $\gamma=1/2$. The left inset in the spectrograms shows the cluster class of the spectra at a given time-step: blue and red colours correspond to blue and red grains; light blue and light red colours correspond to clusters where the \caII lines show weaker emission.}
    \label{fig:temporal_clustermap}
\end{figure*}

Until now, we have only looked at the spectra as individual pixels. We wanted to study the temporal evolution of chromospheric heating events, which could give additional context necessary to discern heating processes. Figure~\ref{fig:cluster_map} shows images in \caII H from snapshot 931 in simulation A. We highlighted the pixels classified as bright blue grains in blue, and pixels classified as bright red grains in red. The blue pixels account for 3.0\% of the total area, and the red pixels for 1.3\%. The cyan and magenta crosses mark the locations that are plotted as spectrograms in Fig.~\ref{fig:temporal_clustermap}. The spectrograms in Fig.~\ref{fig:temporal_clustermap} have cluster information plotted to the left. Here, the dark blue and dark red colours correspond to the same clusters as for the top row (i.e. the bright blue and red grains). The lighter colours indicate clusters with weaker emission in the red or blue part of the calcium lines. A light blue colour is given to columns in clusters 3, 6, 7, 8, 11, 12, 18, and 21; and a light red colour to clusters 4, 5, 9, and 13. White means that the profiles belong to one of the clusters where the calcium lines are in absorption. 

In the \caII H blue wing of Fig.~\ref{fig:cluster_map}, we see the typical bright blue grains marked with blue. The spectrogram of the bright blue grain that is marked in cyan (first row of Fig.~\ref{fig:temporal_clustermap}) shows an oscillating line core with the typical saw-tooth emission feature of upward-propagating shock waves. To the right, we see the integrated chromospheric heating versus time for the same pixel, which rises to $\approx\qty{30}{\kilo W.m^{-2}}$ in total. 

The spectrogram of the magenta pixel is not associated with a blue grain, neither spatially nor temporally. This is shown in the bottom row of the figure, where we can see the time evolution of the pixel marked with the magenta cross. The \caII spectra are formed in a region where there is a shear in horizontal flow that creates strong heating that rises to $\approx\qty{60}{\kilo W.m^{-2}}$ in total. 

\begin{table}
    \centering
    \caption{Average properties of cluster groups.}
    \begin{tabular}{l r r} 
    \hline\hline
    Cluster group      & red grain & blue grain \\ 
    \hline
    area (\unit{\kilo\metre^2}) &  $18\,338$     & $35\,805$ \\
    area (pixels)               &  $4.7$       & $9.2$ \\
    lifetime  (s)               &  $29$        & $42$  \\

    \hline
    \end{tabular}

    \label{tab:cluster_dimensions}
\end{table}

For the pixel marked with a cyan cross in Fig.~\ref{fig:cluster_map}, the $\tau = 1$ height is lower in the blue part of the line core of \caII 8542~\AA, and higher in the red part of the line core of \caII 8542~\AA. 
The emission in the blue wing is formed in an upward-propagating shock front, similar to the line-of-sight velocity seen in the cyan line for cluster 24 in Fig.~\ref{fig:cluster_atmos_blue}. For the spectra marked with the magenta cross, the formation of the red emission peak happens at \qty{0.7}{\mega\metre}. At this height, there is a down-flow in the line-of-sight velocity, and the heating that creates the emission is produced by a gradient in the horizontal velocity component $v_y$ (shear flow). 

\begin{figure*}
    \includegraphics[width=17cm]{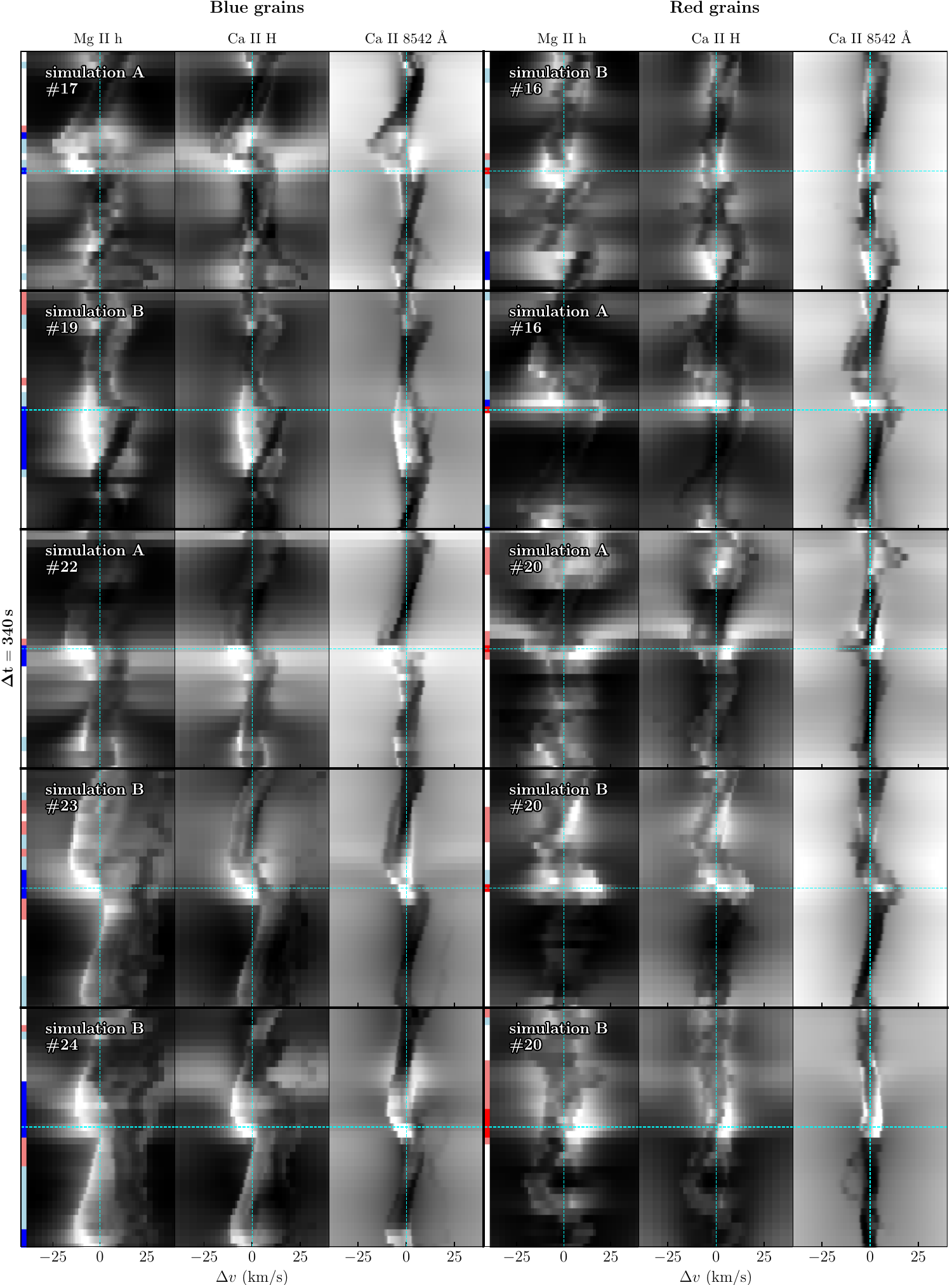}
    \caption{Evolution of spectrograms in the clusters classified as blue grains (left column) and red grains (right column). The horizontal dashed lines mark the snapshot that correspond to the cluster label given in the upper-left corner of each panel, and the vertical dashed lines mark the line centre. The spectra are scaled non-linearly with a $\gamma$ factor to enhance contrast. For \mgII h and \caII H, $\gamma = 1/3$; for \caII 8542~\AA, $\gamma=2/3$. All spectra are plotted with a time interval of $\Delta t = \qty{340}{s}$. The colours in the left inset of the spectrograms show the cluster classes similarly to Fig.~\ref{fig:temporal_clustermap}.}
    \label{fig:cluster_spectrograms}
\end{figure*}

To further investigate the spectral signatures of blue and red grains, we looked at a bigger sample of spectrograms from both of the simulations. We show a subset of spectrograms from clusters labelled as blue and red grains in Fig.~\ref{fig:cluster_spectrograms}. The figure is representative for the dataset as a whole: pixels labelled as bright blue grains usually show a temporal evolution of an oscillating line core with emission in the blue wing as a shock front propagates into the chromosphere. This signature is usually, but not always, present in all three lines. The figure does not show a temporal coherence for bright red grains, something we did not find in the rest of the spectra either. Some red grains show oscillating line cores, while others have no movement in the line core.

Often, pixels were grouped into clusters that were the same or similar as neighbouring pixels. In the colour-coded maps to the left of the spectrograms in Fig.~\ref{fig:cluster_spectrograms}, we plotted the classification of the spectra at each timestep. From these maps, we can see that most blue grains are temporally connected to other blue grains, while red grains usually appear alone. We extended this analysis to all spectra in the dataset by grouping the pixels of bright blue or bright red grains together, and calculated the typical lifetimes and areas of these two groups of clusters. The dimensions of these groups were computed by connecting them in both time and space with bi-connectivity, and discarding cluster groups with fewer than two neighbouring pixels in space. The average lifetime and area of these groups are reported in Table~\ref{tab:cluster_dimensions}. The findings support the trend we see in Fig.~\ref{fig:cluster_spectrograms}, where blue grains have longer lifetimes than red grains. Blue grains also have bigger areas, which is seen in Fig.~\ref{fig:cluster_map} where the red grains seem more scattered.

\section{Discussion}

Using synthetic spectra from two quiet Sun 3D rMHD simulations, we looked for signatures of chromospheric heating in the \mgII h, \caII H, and \caII 8542~\AA\ lines. We clustered the spectra together and find that profiles associated with the strongest chromospheric heating have strong emission and are mostly of the type associated with chromospheric bright grains, produced by upward-propagating acoustic shocks, as shown in Fig.~\ref{fig:cluster_atmos_blue}. 

Typical quiet Sun spectral observations show three-minute oscillations from $p$-mode waves. Transient brightenings in the \caII lines are signs of heating caused by the steepening of $p$-mode waves. This internetwork region is sometimes called the clapotissphere \citep{Rutten:1991aa,Rutten:1995aa} due to the steep shock waves that characterise this area. Chromospheric shock waves also create oscillations in the spectra with sawtooth shaped emission in the blue wing. Therefore, \caII bright grains have long been a signature of chromospheric wave heating \citep[e.g.][]{Carlsson:1992vg}. 

In the simulations, chromospheric bright grains were relatively rare, but associated with an excess of chromospheric heating. The strong blue emission clusters accounted for only 3.3\% of the pixels in simulation A and 2.6\% of the pixels in simulation B. However, these pixels had heating rates much larger than the average: four times higher in simulation A and three times higher in simulation B. Bright blue grains dominated the heating locally, and their fraction of the total chromospheric heating balance was as high as 13.4\% in simulation A, and 8.1\% in simulation B. Using a similar simulation but different methods, \citet{Noraz:2025aa} found that 15\% of chromospheric energy dissipation comes from shocks, which is comparable, but slightly higher, than our results. Not all shock-front-heating in our simulations was related to a bright blue grain spectral signature. For instance, some shock front heating happened beneath the height of formation of the \caII 8542~\AA~line. In these cases, as the hot plasma was advected upwards in the chromosphere, emission in the \caII and \mgII lines was seen after one time-step (\qty{10}{s}). Therefore, our estimation of shock-front heating from blue grains is likely to undershoot by a few percent if we compare with the results of \citet{Noraz:2025aa}.

Simulation A had about 40\% more bright grains than simulation B. However, in both simulations the strong heating was confined to a small area (50\% of the heating came from 15\% of the area as seen in Fig.~\ref{fig:cumulative_heating}). Therefore, a larger proportion of the chromospheric heating in simulation B does not show clear spectral signatures in the \caII and \mgII lines. Since the main difference between these simulations is the magnetic configuration, this suggests that the inclination of the chromospheric magnetic field determines the occurrence of bright grains.

Other spectral profiles associated with large chromospheric heating have strong emission in the red wing. These were included in two groups of clusters (16 and 20). Unlike the bright blue grain profiles, these profiles did not share a common atmospheric profile nor temporal evolution. Instead, they appear to be formed in a variety of conditions. The profiles in cluster 16 were produced by two different groups of atmospheres. Most of the emission profiles were formed in the lower chromosphere in regions with very low velocities, with an up-flow above that makes the atmosphere transparent to brightenings in the red wing. A smaller fraction of the profiles were actually formed in chromospheric down-flows. Emission in the red wing of the calcium lines can also be caused by magnetic flux emergence \citep[magnetic bubbles,][]{Ortiz:2014aa,de-la-Cruz-Rodriguez:2015aa}, but in this cluster we saw no conclusive trends in the magnetic field. \citet{Ortiz:2014aa} also explain red emission with enhanced absorption in the blue wing, which is also the case for the up-flowing atmospheres we saw in cluster 16, where the blue wing was formed higher in the chromosphere. The degeneracy in the formation of the red emission peak in calcium lines in clusters 16 and 20 makes it important to use caution when analysing such profiles from observations, as seen by the ambiguity in velocities inferred from \caII 8542~\AA~alone \citep{Kuckein:2025aa}. A possible way to resolve the degeneracy of different heating profiles would be to use additional spectral lines, or use different Stokes components to resolve the degeneracy of different heating processes (e.g. as \citealt{Zhou:2025aa} have done to identify signatures of reconnection in the 8542~\AA\ line).

The bright red grains comprised only 0.7\% of all profiles, but similar to the bright blue grains, their chromospheric heating was much larger than the average (four times higher for simulation A and 3.6 times higher for simulation B). Their contribution to the total chromospheric heating is 3.3\% for simulation A and 2.2\% for simulation B. This is lower than the contribution of the blue grains because red grains are not as common. If we include all clusters where \caII H has emission in the line core, we obtain a much larger proportion of the total heating. In simulation A these pixels make up 47\% of the total chromospheric heating, and in simulation B the same proportion is 38\%.

In absolute values, the spatially and temporally averaged chromospheric heating in our simulation was \qty{3.8}{\kilo W.\metre^{-2}} for simulation A and \qty{6.2}{\kilo W.\metre^{-2}} for simulation B. These values are comparable to the canonical value of radiative losses in the VAL chromosphere of \qty{4.6}{\kilo W.\metre^{-2}} \citep{Vernazza:1981aa}. The average heating from blue and bright grains was substantially higher: \qty{15.2}{\kilo W.\metre^{-2}} and \qty{19.4}{\kilo W.\metre^{-2}} for the blue grains; and \qty{15.2}{\kilo W.\metre^{-2}} and \qty{22.4}{\kilo W.\metre^{-2}} for the red grains. Given that bright grains are rare and their heating is far higher than the average values, it is best to compare them to spatially resolved heating estimates. \citet{Diaz-Baso:2021aa} make use of inversions of solar observations to derive spatially resolved estimates of radiative losses in an active region. Averaging over small patches, they find radiative losses of \qty{4.5}{\kilo W.\metre^{-2}} for the quiet Sun, \qty{32}{\kilo W.\metre^{-2}} for a weaker plage, and \qty{108}{\kilo W.\metre^{-2}} for the most active patches. \citet{Morosin:2022aa} use similar methods and find average radiative losses of \qty{28}{\kilo W.\metre^{-2}} for a plage. These plage regions have magnetic fields and conditions much more extreme than our simulated setup, but both \citet{Diaz-Baso:2021aa} and \citet{Morosin:2022aa} find many locations farther from the most active parts that locally have radiative losses in the range of 10--\qty{20}{\kilo W.\metre^{-2}}, fully consistent with our values for the bright grains.

The temperature excursions in shock fronts that produce the bright blue grain signatures were found to be around \qty{2500}{K} on average. This value is comparable to those inferred from the inversion of SST observations \citep[\qtyrange{1000}{2000}{K} in][]{de-la-Cruz-Rodriguez:2013aa,Joshi:2018aa,Mathur:2022aa}, but somewhat larger than values inferred from brightness temperature at millimetre wavelengths \citep[\qtyrange{450}{750}{K}, see][]{Eklund:2020aa}. The ALMA observations of \citet{Eklund:2020aa} have a spatial resolution of about 2\farcs0, which can very well average small-scale temperature peaks. From Fig.~\ref{fig:cluster_map} we see that strong emission regions appear at scales smaller than $\approx$250~km, or about 0\farcs3.

The average lifetime of blue emission signatures was almost one minute, consistent with bright grains observed in \caII 8542~\AA~\citep{Martinez-Gonzalez:2023aa} and \caII H2V \citep{Mathur:2022aa}. The average area of blue grains corresponds to an angular diameter of $0\farcs3$, and of $0\farcs2$ for the red grains. The bright grains in our simulations are smaller than previously reported from observations, with \caII 8542~\AA~grain sizes seen with IBIS ranging from $0\farcs7$ -- $1\farcs0$ \citep{Vecchio:2009aa}. This can be due to the higher spatial resolution of the synthetic data.

Our analysis is limited to the heating that occurs in these simulations. Bifrost uses a hyper-diffusion scheme, which makes the code stable for a broad range of different plasma regimes. However, this scheme smooths out gradients and spreads out heating, which should make heating from shock fronts a lower bound \citep{Udnaes:2025aa}. Viscosity and resistivity have a large impact on the plasma dynamics. \citet{Faerder:2024aa} study the impact of different resistivity models in Bifrost, which is important for the nature of reconnection and heating in the simulations. They find that the hyper-diffusive model currently employed in Bifrost best reproduced the total heating at lower resolutions. Therefore, while our simulations have different resistivities and viscosities than the solar atmosphere, their approximations to chromospheric heating is as close as we can get with current methods.

\section{Conclusions}

Chromospheric heating is complex and cannot be measured directly. The key spectral diagnostics we can observe form under optically thick non-LTE conditions, whose non-locality can lead to loss of information. In this study, we use 3D rMHD simulations combined with spatio-temporal non-LTE spectral synthesis to understand how to quantify heating (which we can measure directly in the simulations) from spectral signatures.

We calculated chromospheric spectra in non-LTE from two quiet Sun rMHD simulations with different magnetic configurations: one typical quiet Sun simulation (A), and one simulation with a coronal hole-like magnetic field (B). By using k-means clustering, we identified heating signatures in the chromospheric spectra. The spectral signatures associated with the strongest heating showed strong emission on either the blue or red wings of all lines, and we refer to them as blue or red grains, respectively. 

We find that blue grains were produced by upward-propagating shock waves, while red grains does not show a typical atmospheric stratification. The blue grains have a coherent temporal evolution with an oscillating line core and sawtooth-like emission in the blue wing. Red grains have many different temporal signatures, including oscillating line cores with red wing emission and stationary line cores with sudden emission in the inner red wing.

The clusters with strong emission accounted for fewer than 4\% of all the profiles in the dataset, but these profiles represented 17\% of the chromospheric energy dissipation in simulation A, and 10\% of the chromospheric energy dissipation in simulation B. There were 40\% more bright grains in simulation A compared to simulation B, reflecting a dependence on the magnetic-field orientation, which was unipolar and vertical in simulation B.

Including all clusters with emission in the \caII~H line accounted for 47\% and 38\% of the heating in simulation A and B, respectively. We find strong emission to be relatively rare in our simulations, but dominating energy dissipation locally. In the simulations, a substantial portion of the chromospheric heating is dynamic with lifetimes under a minute, and localised in regions of a few \qty{1e4}{\km^2}.

\begin{acknowledgements}
This work has been supported by the Research Council of Norway through its Centres of Excellence scheme, project number 262622. We kindly acknowledge the computational resources provided by UNINETT Sigma2 - the National Infrastructure for High Performance Computing and Data Storage in Norway.
\end{acknowledgements}

\bibpunct{(}{)}{;}{a}{}{,}
\bibliographystyle{aa}
\bibliography{references}

\begin{thebibliography}{67}
\expandafter\ifx\csname natexlab\endcsname\relax\def\natexlab#1{#1}\fi

\bibitem[{{Abbasvand} {et~al.}(2021){Abbasvand}, {Sobotka}, {{\v{S}}vanda}, {Heinzel}, {Liu}, \& {Mravcov{\'a}}}]{Abbasvand:2021aa}
{Abbasvand}, V., {Sobotka}, M., {{\v{S}}vanda}, M., {et~al.} 2021, \aap, 648, A28

\bibitem[{{Asplund} {et~al.}(2009){Asplund}, {Grevesse}, {Sauval}, \& {Scott}}]{Asplund:2009aa}
{Asplund}, M., {Grevesse}, N., {Sauval}, A.~J., \& {Scott}, P. 2009, \araa, 47, 481

\bibitem[{{Barczynski} {et~al.}(2021){Barczynski}, {Harra}, {Kleint}, {Panos}, \& {Brooks}}]{Barczynski:2021aa}
{Barczynski}, K., {Harra}, L., {Kleint}, L., {Panos}, B., \& {Brooks}, D.~H. 2021, \aap, 651, A112

\bibitem[{{Beckers}(1964)}]{Beckers:1964aa}
{Beckers}, J.~M. 1964, PhD thesis, University of Utrecht, Netherlands

\bibitem[{{Bello Gonz{\'a}lez} {et~al.}(2009){Bello Gonz{\'a}lez}, {Flores Soriano}, {Kneer}, \& {Okunev}}]{Bello-Gonzalez:2009aa}
{Bello Gonz{\'a}lez}, N., {Flores Soriano}, M., {Kneer}, F., \& {Okunev}, O. 2009, \aap, 508, 941

\bibitem[{{Bj{\o}rgen} {et~al.}(2018){Bj{\o}rgen}, {Sukhorukov}, {Leenaarts}, {Carlsson}, {de la Cruz Rodr{\'\i}guez}, {Scharmer}, \& {Hansteen}}]{Bjorgen:2018aa}
{Bj{\o}rgen}, J.~P., {Sukhorukov}, A.~V., {Leenaarts}, J., {et~al.} 2018, \aap, 611, A62

\bibitem[{{Brchnelova} {et~al.}(2025){Brchnelova}, {Gudiksen}, {Carlsson}, {Lani}, \& {Poedts}}]{Brchnelova:2025aa}
{Brchnelova}, M., {Gudiksen}, B., {Carlsson}, M., {Lani}, A., \& {Poedts}, S. 2025, \aap, 693, A74

\bibitem[{{Carlsson} {et~al.}(2019){Carlsson}, {De Pontieu}, \& {Hansteen}}]{Carlsson:2019aa}
{Carlsson}, M., {De Pontieu}, B., \& {Hansteen}, V.~H. 2019, \araa, 57, 189

\bibitem[{{Carlsson} \& {Leenaarts}(2012)}]{Carlsson:2012aa}
{Carlsson}, M. \& {Leenaarts}, J. 2012, \aap, 539, A39

\bibitem[{{Carlsson} \& {Stein}(1992)}]{Carlsson:1992vg}
{Carlsson}, M. \& {Stein}, R.~F. 1992, \apjl, 397, L59

\bibitem[{{da Silva Santos} {et~al.}(2024){da Silva Santos}, {Molnar}, {Mili{\'c}}, {Rempel}, {Reardon}, \& {de la Cruz Rodr{\'\i}guez}}]{da-Silva-Santos:2024aa}
{da Silva Santos}, J.~M., {Molnar}, M., {Mili{\'c}}, I., {et~al.} 2024, \apj, 976, 21

\bibitem[{{de la Cruz Rodr{\'\i}guez} {et~al.}(2015){de la Cruz Rodr{\'\i}guez}, {Hansteen}, {Bellot-Rubio}, \& {Ortiz}}]{de-la-Cruz-Rodriguez:2015aa}
{de la Cruz Rodr{\'\i}guez}, J., {Hansteen}, V., {Bellot-Rubio}, L., \& {Ortiz}, A. 2015, \apj, 810, 145

\bibitem[{{de la Cruz Rodr{\'\i}guez} {et~al.}(2013){de la Cruz Rodr{\'\i}guez}, {Rouppe van der Voort}, {Socas-Navarro}, \& {van Noort}}]{de-la-Cruz-Rodriguez:2013aa}
{de la Cruz Rodr{\'\i}guez}, J., {Rouppe van der Voort}, L., {Socas-Navarro}, H., \& {van Noort}, M. 2013, \aap, 556, A115

\bibitem[{{de la Cruz Rodr{\'\i}guez} {et~al.}(2012){de la Cruz Rodr{\'\i}guez}, {Socas-Navarro}, {Carlsson}, \& {Leenaarts}}]{de-la-Cruz-Rodriguez:2012aa}
{de la Cruz Rodr{\'\i}guez}, J., {Socas-Navarro}, H., {Carlsson}, M., \& {Leenaarts}, J. 2012, \aap, 543, A34

\bibitem[{{De Pontieu} {et~al.}(2015){De Pontieu}, {McIntosh}, {Martinez-Sykora}, {Peter}, \& {Pereira}}]{De-Pontieu:2015aa}
{De Pontieu}, B., {McIntosh}, S., {Martinez-Sykora}, J., {Peter}, H., \& {Pereira}, T.~M.~D. 2015, \apjl, 799, L12

\bibitem[{{De Pontieu} {et~al.}(2014){De Pontieu}, {Title}, {Lemen}, {Kushner}, {Akin}, {Allard}, {Berger}, {Boerner}, {Cheung}, {Chou}, {Drake}, {Duncan}, {Freeland}, {Heyman}, {Hoffman}, {Hurlburt}, {Lindgren}, {Mathur}, {Rehse}, {Sabolish}, {Seguin}, {Schrijver}, {Tarbell}, {W{\"u}lser}, {Wolfson}, {Yanari}, {Mudge}, {Nguyen-Phuc}, {Timmons}, {van Bezooijen}, {Weingrod}, {Brookner}, {Butcher}, {Dougherty}, {Eder}, {Knagenhjelm}, {Larsen}, {Mansir}, {Phan}, {Boyle}, {Cheimets}, {DeLuca}, {Golub}, {Gates}, {Hertz}, {McKillop}, {Park}, {Perry}, {Podgorski}, {Reeves}, {Saar}, {Testa}, {Tian}, {Weber}, {Dunn}, {Eccles}, {Jaeggli}, {Kankelborg}, {Mashburn}, {Pust}, {Springer}, {Carvalho}, {Kleint}, {Marmie}, {Mazmanian}, {Pereira}, {Sawyer}, {Strong}, {Worden}, {Carlsson}, {Hansteen}, {Leenaarts}, {Wiesmann}, {Aloise}, {Chu}, {Bush}, {Scherrer}, {Brekke}, {Martinez-Sykora}, {Lites}, {McIntosh}, {Uitenbroek}, {Okamoto}, {Gummin}, {Auker}, {Jerram}, {Pool}, \& {Waltham}}]{De-Pontieu:2014aa}
{De Pontieu}, B., {Title}, A.~M., {Lemen}, J.~R., {et~al.} 2014, \solphys, 289, 2733

\bibitem[{{D{\'\i}az Baso} {et~al.}(2021){D{\'\i}az Baso}, {de la Cruz Rodr{\'\i}guez}, \& {Leenaarts}}]{Diaz-Baso:2021aa}
{D{\'\i}az Baso}, C.~J., {de la Cruz Rodr{\'\i}guez}, J., \& {Leenaarts}, J. 2021, \aap, 647, A188

\bibitem[{{Eklund} {et~al.}(2020){Eklund}, {Wedemeyer}, {Szydlarski}, {Jafarzadeh}, \& {Guevara G{\'o}mez}}]{Eklund:2020aa}
{Eklund}, H., {Wedemeyer}, S., {Szydlarski}, M., {Jafarzadeh}, S., \& {Guevara G{\'o}mez}, J.~C. 2020, \aap, 644, A152

\bibitem[{{F{\ae}rder} {et~al.}(2024){F{\ae}rder}, {N{\'o}brega-Siverio}, \& {Carlsson}}]{Faerder:2024aa}
{F{\ae}rder}, {\O}.~H., {N{\'o}brega-Siverio}, D., \& {Carlsson}, M. 2024, \aap, 683, A95

\bibitem[{{Finley} {et~al.}(2022){Finley}, {Brun}, {Carlsson}, {Szydlarski}, {Hansteen}, \& {Shoda}}]{Finley:2022aa}
{Finley}, A.~J., {Brun}, A.~S., {Carlsson}, M., {et~al.} 2022, \aap, 665, A118

\bibitem[{{Fossum} \& {Carlsson}(2005)}]{Fossum:2005aa}
{Fossum}, A. \& {Carlsson}, M. 2005, \nat, 435, 919

\bibitem[{{Fossum} \& {Carlsson}(2006)}]{Fossum:2006aa}
{Fossum}, A. \& {Carlsson}, M. 2006, \apj, 646, 579

\bibitem[{{Freytag} {et~al.}(2012){Freytag}, {Steffen}, {Ludwig}, {Wedemeyer-B{\"o}hm}, {Schaffenberger}, \& {Steiner}}]{Freytag:2012aa}
{Freytag}, B., {Steffen}, M., {Ludwig}, H.~G., {et~al.} 2012, Journal of Computational Physics, 231, 919

\bibitem[{{Gudiksen} \& {Nordlund}(2005)}]{Gudiksen:2005aa}
{Gudiksen}, B.~V. \& {Nordlund}, {\r{A}}. 2005, \apj, 618, 1020

\bibitem[{{Hansteen} {et~al.}(2023){Hansteen}, {Martinez-Sykora}, {Carlsson}, {De Pontieu}, {Go{\v{s}}i{\'c}}, \& {Bose}}]{Hansteen:2023aa}
{Hansteen}, V.~H., {Martinez-Sykora}, J., {Carlsson}, M., {et~al.} 2023, \apj, 944, 131

\bibitem[{{Joshi} \& {de la Cruz Rodr{\'\i}guez}(2018)}]{Joshi:2018aa}
{Joshi}, J. \& {de la Cruz Rodr{\'\i}guez}, J. 2018, \aap, 619, A63

\bibitem[{{Joshi} {et~al.}(2020){Joshi}, {Rouppe van der Voort}, \& {de la Cruz Rodr{\'\i}guez}}]{Joshi:2020aa}
{Joshi}, J., {Rouppe van der Voort}, L. H.~M., \& {de la Cruz Rodr{\'\i}guez}, J. 2020, \aap, 641, L5

\bibitem[{{Khomenko} {et~al.}(2003){Khomenko}, {Collados}, {Solanki}, {Lagg}, \& {Trujillo Bueno}}]{Khomenko:2003aa}
{Khomenko}, E.~V., {Collados}, M., {Solanki}, S.~K., {Lagg}, A., \& {Trujillo Bueno}, J. 2003, \aap, 408, 1115

\bibitem[{{Kleint} \& {Panos}(2022)}]{Kleint:2022aa}
{Kleint}, L. \& {Panos}, B. 2022, \aap, 657, A132

\bibitem[{{Kuckein} {et~al.}(2025){Kuckein}, {Collados}, {Asensio Ramos}, {D{\'\i}az Baso}, {Felipe}, {Quintero Noda}, {Kleint}, {Fletcher}, \& {Matthews}}]{Kuckein:2025aa}
{Kuckein}, C., {Collados}, M., {Asensio Ramos}, A., {et~al.} 2025, \aap, 699, A121

\bibitem[{{Kuridze} {et~al.}(2017){Kuridze}, {Henriques}, {Mathioudakis}, {Koza}, {Zaqarashvili}, {Ryb{\'a}k}, {Hanslmeier}, \& {Keenan}}]{Kuridze:2017aa}
{Kuridze}, D., {Henriques}, V., {Mathioudakis}, M., {et~al.} 2017, \apj, 846, 9

\bibitem[{{Leenaarts} {et~al.}(2009){Leenaarts}, {Carlsson}, {Hansteen}, \& {Rouppe van der Voort}}]{Leenaarts:2009aa}
{Leenaarts}, J., {Carlsson}, M., {Hansteen}, V., \& {Rouppe van der Voort}, L. 2009, \apjl, 694, L128

\bibitem[{{Leenaarts} {et~al.}(2013{\natexlab{a}}){Leenaarts}, {Pereira}, {Carlsson}, {Uitenbroek}, \& {De Pontieu}}]{Leenaarts:2013ab}
{Leenaarts}, J., {Pereira}, T.~M.~D., {Carlsson}, M., {Uitenbroek}, H., \& {De Pontieu}, B. 2013{\natexlab{a}}, \apj, 772, 89

\bibitem[{{Leenaarts} {et~al.}(2013{\natexlab{b}}){Leenaarts}, {Pereira}, {Carlsson}, {Uitenbroek}, \& {De Pontieu}}]{Leenaarts:2013aa}
{Leenaarts}, J., {Pereira}, T.~M.~D., {Carlsson}, M., {Uitenbroek}, H., \& {De Pontieu}, B. 2013{\natexlab{b}}, \apj, 772, 90

\bibitem[{{Mart{\'\i}nez Gonz{\'a}lez} {et~al.}(2023){Mart{\'\i}nez Gonz{\'a}lez}, {del Pino Alem{\'a}n}, {Pastor Yabar}, {Quintero Noda}, \& {Asensio Ramos}}]{Martinez-Gonzalez:2023aa}
{Mart{\'\i}nez Gonz{\'a}lez}, M.~J., {del Pino Alem{\'a}n}, T., {Pastor Yabar}, A., {Quintero Noda}, C., \& {Asensio Ramos}, A. 2023, \apjl, 955, L40

\bibitem[{{Mathur} {et~al.}(2025){Mathur}, {Anusha}, \& {Agnihotri}}]{Mathur:2025aa}
{Mathur}, H., {Anusha}, L.~S., \& {Agnihotri}, D. 2025, \apj, 983, 84

\bibitem[{{Mathur} {et~al.}(2022){Mathur}, {Joshi}, {Nagaraju}, {Rouppe van der Voort}, \& {Bose}}]{Mathur:2022aa}
{Mathur}, H., {Joshi}, J., {Nagaraju}, K., {Rouppe van der Voort}, L., \& {Bose}, S. 2022, \aap, 668, A153

\bibitem[{{Moe} {et~al.}(2023){Moe}, {Pereira}, {Calvo}, \& {Leenaarts}}]{Moe:2023aa}
{Moe}, T.~E., {Pereira}, T. M.~D., {Calvo}, F., \& {Leenaarts}, J. 2023, \aap, 675, A130

\bibitem[{{Moe} {et~al.}(2022){Moe}, {Pereira}, \& {Carlsson}}]{Moe:2022aa}
{Moe}, T.~E., {Pereira}, T. M.~D., \& {Carlsson}, M. 2022, \aap, 662, A80

\bibitem[{{Moe} {et~al.}(2024){Moe}, {Pereira}, {Rouppe van der Voort}, {Carlsson}, {Hansteen}, {Calvo}, \& {Leenaarts}}]{Moe:2024aa}
{Moe}, T.~E., {Pereira}, T. M.~D., {Rouppe van der Voort}, L., {et~al.} 2024, \aap, 682, A11

\bibitem[{{Molnar} {et~al.}(2023){Molnar}, {Reardon}, {Cranmer}, {Kowalski}, \& {Mili{\'c}}}]{Molnar:2023aa}
{Molnar}, M.~E., {Reardon}, K.~P., {Cranmer}, S.~R., {Kowalski}, A.~F., \& {Mili{\'c}}, I. 2023, \apj, 945, 154

\bibitem[{{Morosin} {et~al.}(2022){Morosin}, {de la Cruz Rodr{\'\i}guez}, {D{\'\i}az Baso}, \& {Leenaarts}}]{Morosin:2022aa}
{Morosin}, R., {de la Cruz Rodr{\'\i}guez}, J., {D{\'\i}az Baso}, C.~J., \& {Leenaarts}, J. 2022, \aap, 664, A8

\bibitem[{{N{\'o}brega-Siverio} {et~al.}(2021){N{\'o}brega-Siverio}, {Guglielmino}, \& {Sainz Dalda}}]{Nobrega-Siverio:2021aa}
{N{\'o}brega-Siverio}, D., {Guglielmino}, S.~L., \& {Sainz Dalda}, A. 2021, \aap, 655, A28

\bibitem[{{Noraz} {et~al.}(2025){Noraz}, {Carlsson}, \& {Aulanier}}]{Noraz:2025aa}
{Noraz}, Q., {Carlsson}, M., \& {Aulanier}, G. 2025, arXiv e-prints, arXiv:2511.01858

\bibitem[{{Ondratschek} {et~al.}(2024){Ondratschek}, {Przybylski}, {Smitha}, {Cameron}, {Solanki}, \& {Leenaarts}}]{Ondratschek:2024aa}
{Ondratschek}, P., {Przybylski}, D., {Smitha}, H.~N., {et~al.} 2024, \aap, 692, A6

\bibitem[{{Ortiz} {et~al.}(2014){Ortiz}, {Bellot Rubio}, {Hansteen}, {de la Cruz Rodr{\'\i}guez}, \& {Rouppe van der Voort}}]{Ortiz:2014aa}
{Ortiz}, A., {Bellot Rubio}, L.~R., {Hansteen}, V.~H., {de la Cruz Rodr{\'\i}guez}, J., \& {Rouppe van der Voort}, L. 2014, \apj, 781, 126

\bibitem[{{Pereira} {et~al.}(2013{\natexlab{a}}){Pereira}, {Asplund}, {Collet}, {Thaler}, {Trampedach}, \& {Leenaarts}}]{Pereira:2013ab}
{Pereira}, T.~M.~D., {Asplund}, M., {Collet}, R., {et~al.} 2013{\natexlab{a}}, \aap, 554, A118

\bibitem[{{Pereira} {et~al.}(2013{\natexlab{b}}){Pereira}, {Leenaarts}, {De Pontieu}, {Carlsson}, \& {Uitenbroek}}]{Pereira:2013aa}
{Pereira}, T.~M.~D., {Leenaarts}, J., {De Pontieu}, B., {Carlsson}, M., \& {Uitenbroek}, H. 2013{\natexlab{b}}, \apj, 778, 143

\bibitem[{{Pereira} \& {Uitenbroek}(2015)}]{Pereira:2015wv}
{Pereira}, T. M.~D. \& {Uitenbroek}, H. 2015, \aap, 574, A3

\bibitem[{{Pietarila} {et~al.}(2007){Pietarila}, {Socas-Navarro}, \& {Bogdan}}]{Pietarila:2007aa}
{Pietarila}, A., {Socas-Navarro}, H., \& {Bogdan}, T. 2007, \apj, 663, 1386

\bibitem[{{Przybylski} {et~al.}(2022){Przybylski}, {Cameron}, {Solanki}, {Rempel}, {Leenaarts}, {Anusha}, {Witzke}, \& {Shapiro}}]{Przybylski:2022aa}
{Przybylski}, D., {Cameron}, R., {Solanki}, S.~K., {et~al.} 2022, \aap, 664, A91

\bibitem[{{Quintero Noda} {et~al.}(2016){Quintero Noda}, {Shimizu}, {de la Cruz Rodr{\'\i}guez}, {Katsukawa}, {Ichimoto}, {Anan}, \& {Suematsu}}]{Quintero-Noda:2016aa}
{Quintero Noda}, C., {Shimizu}, T., {de la Cruz Rodr{\'\i}guez}, J., {et~al.} 2016, \mnras, 459, 3363

\bibitem[{{Rutten}(1995)}]{Rutten:1995aa}
{Rutten}, R.~J. 1995, in ESA Special Publication, Vol. 376, Helioseismology, ed. J.~T. {Hoeksema}, V.~{Domingo}, B.~{Fleck}, \& B.~{Battrick}, 151

\bibitem[{{Rutten} \& {Uitenbroek}(1991)}]{Rutten:1991aa}
{Rutten}, R.~J. \& {Uitenbroek}, H. 1991, \solphys, 134, 15

\bibitem[{{Scharmer} {et~al.}(2003){Scharmer}, {Bjelksjo}, {Korhonen}, {Lindberg}, \& {Petterson}}]{Scharmer:2003aa}
{Scharmer}, G.~B., {Bjelksjo}, K., {Korhonen}, T.~K., {Lindberg}, B., \& {Petterson}, B. 2003, in Society of Photo-Optical Instrumentation Engineers (SPIE) Conference Series, Vol. 4853, Innovative Telescopes and Instrumentation for Solar Astrophysics, ed. S.~L. {Keil} \& S.~V. {Avakyan}, 341--350

\bibitem[{{Silva} {et~al.}(2022){Silva}, {Murabito}, {Jafarzadeh}, {Stangalini}, {Verth}, {Ballai}, \& {Fedun}}]{Silva:2022aa}
{Silva}, S. S.~A., {Murabito}, M., {Jafarzadeh}, S., {et~al.} 2022, \apj, 927, 146

\bibitem[{{Udn{\ae}s} \& {Pereira}(2025)}]{Udnaes:2025aa}
{Udn{\ae}s}, E.~R. \& {Pereira}, T. M.~D. 2025, \aap, 699, A25

\bibitem[{{Uitenbroek}(2001)}]{Uitenbroek:2001wf}
{Uitenbroek}, H. 2001, \apj, 557, 389

\bibitem[{{Vecchio} {et~al.}(2009){Vecchio}, {Cauzzi}, \& {Reardon}}]{Vecchio:2009aa}
{Vecchio}, A., {Cauzzi}, G., \& {Reardon}, K.~P. 2009, \aap, 494, 269

\bibitem[{{Vernazza} {et~al.}(1981){Vernazza}, {Avrett}, \& {Loeser}}]{Vernazza:1981aa}
{Vernazza}, J.~E., {Avrett}, E.~H., \& {Loeser}, R. 1981, \apjs, 45, 635

\bibitem[{{Viticchi{\'e}} \& {S{\'a}nchez Almeida}(2011)}]{Viticchie:2011aa}
{Viticchi{\'e}}, B. \& {S{\'a}nchez Almeida}, J. 2011, \aap, 530, A14

\bibitem[{{V{\"o}gler} {et~al.}(2005){V{\"o}gler}, {Shelyag}, {Sch{\"u}ssler}, {Cattaneo}, {Emonet}, \& {Linde}}]{Vogler:2005aa}
{V{\"o}gler}, A., {Shelyag}, S., {Sch{\"u}ssler}, M., {et~al.} 2005, \aap, 429, 335

\bibitem[{{Wedemeyer} {et~al.}(2022){Wedemeyer}, {Fleishman}, {de la Cruz Rodr{\'\i}guez}, {Gun{\'a}r}, {da Silva Santos}, {Antolin}, {Guevara G{\'o}mez}, {Szydlarski}, \& {Eklund}}]{Wedemeyer:2022aa}
{Wedemeyer}, S., {Fleishman}, G., {de la Cruz Rodr{\'\i}guez}, J., {et~al.} 2022, Frontiers in Astronomy and Space Sciences, 9, 335

\bibitem[{{Wedemeyer-B{\"o}hm} \& {Carlsson}(2011)}]{Wedemeyer-Bohm:2011aa}
{Wedemeyer-B{\"o}hm}, S. \& {Carlsson}, M. 2011, \aap, 528, A1

\bibitem[{{Wedemeyer-B{\"o}hm} {et~al.}(2007){Wedemeyer-B{\"o}hm}, {Steiner}, {Bruls}, \& {Rammacher}}]{Wedemeyer-Bohm:2007aa}
{Wedemeyer-B{\"o}hm}, S., {Steiner}, O., {Bruls}, J., \& {Rammacher}, W. 2007, in Astronomical Society of the Pacific Conference Series, Vol. 368, The Physics of Chromospheric Plasmas, ed. P.~{Heinzel}, I.~{Dorotovi{\v{c}}}, \& R.~J. {Rutten}, 93

\bibitem[{{Woods} {et~al.}(2021){Woods}, {Sainz Dalda}, \& {De Pontieu}}]{Woods:2021aa}
{Woods}, M.~M., {Sainz Dalda}, A., \& {De Pontieu}, B. 2021, \apj, 922, 137

\bibitem[{{Zhou} {et~al.}(2025){Zhou}, {Yokoyama}, {Iijima}, {Matsumoto}, {Toriumi}, {Katsukawa}, \& {Kubo}}]{Zhou:2025aa}
{Zhou}, X., {Yokoyama}, T., {Iijima}, H., {et~al.} 2025, \apj, 993, 43

\end{thebibliography}

\begin{appendix}

\section{Heating in the simulations} \label{app:A}

\begin{figure}
    \resizebox{\hsize}{!}{\includegraphics{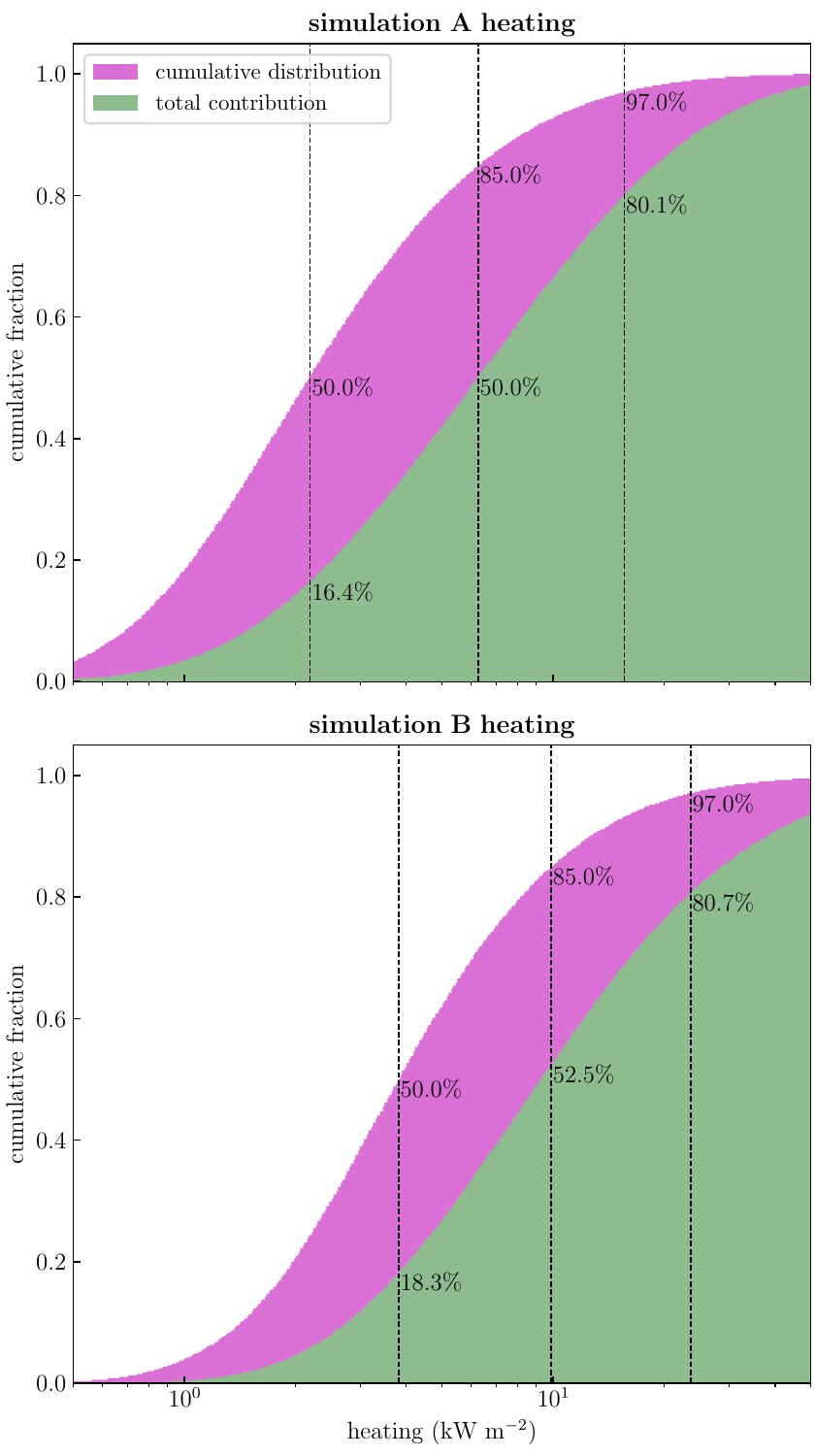}}
    \caption{Cumulative distributions of chromospheric heating and the contribution to the total chromospheric heating. The dashed black lines show percentiles of the two distributions.}
    \label{fig:cumulative_heating}
\end{figure}

In the simulations we studied, the distribution of dissipative chromospheric heating was highly skewed. We show the cumulative distribution and the contribution to dissipative heating in Fig.~\ref{fig:cumulative_heating}. Small areas of the simulations contribute to a large fraction of the total chromospheric heating. E.g. 50\% of the total heating comes from only 15\% of the total area.

\section{K-means clustering} \label{app:B}

In this section, we describe the effects of the pruning of clusters. In general, more clusters are favourable in order to capture more variation in line profiles. But, increasing the number of clusters also increases the model's complexity, bias, and interpretability.

\begin{figure}
    \resizebox{\hsize}{!}{\includegraphics{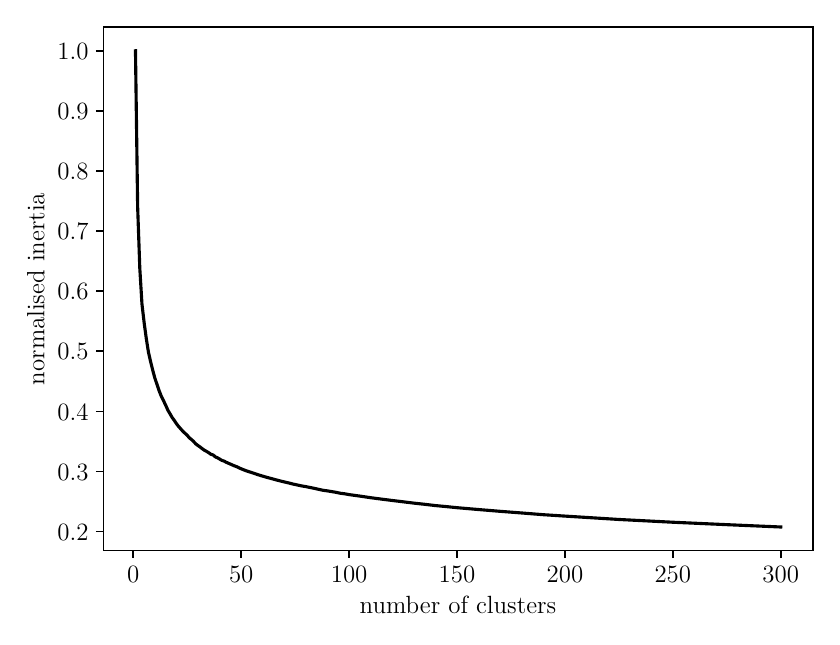}}
    \caption{Inertia versus number of clusters of the k-means clustering on the spectra.}
    \label{fig:inertia}
\end{figure}

A textbook example to determine the optimal number of clusters is the elbow method. We show inertia as a function of clusters in Fig.~\ref{fig:inertia}, and observe that there does not exist an elbow for our data. Therefore, there is no straightforward way to determine the perfect number of clusters for our data. 

\begin{figure*}
    \sidecaption
    \includegraphics[width=12cm]{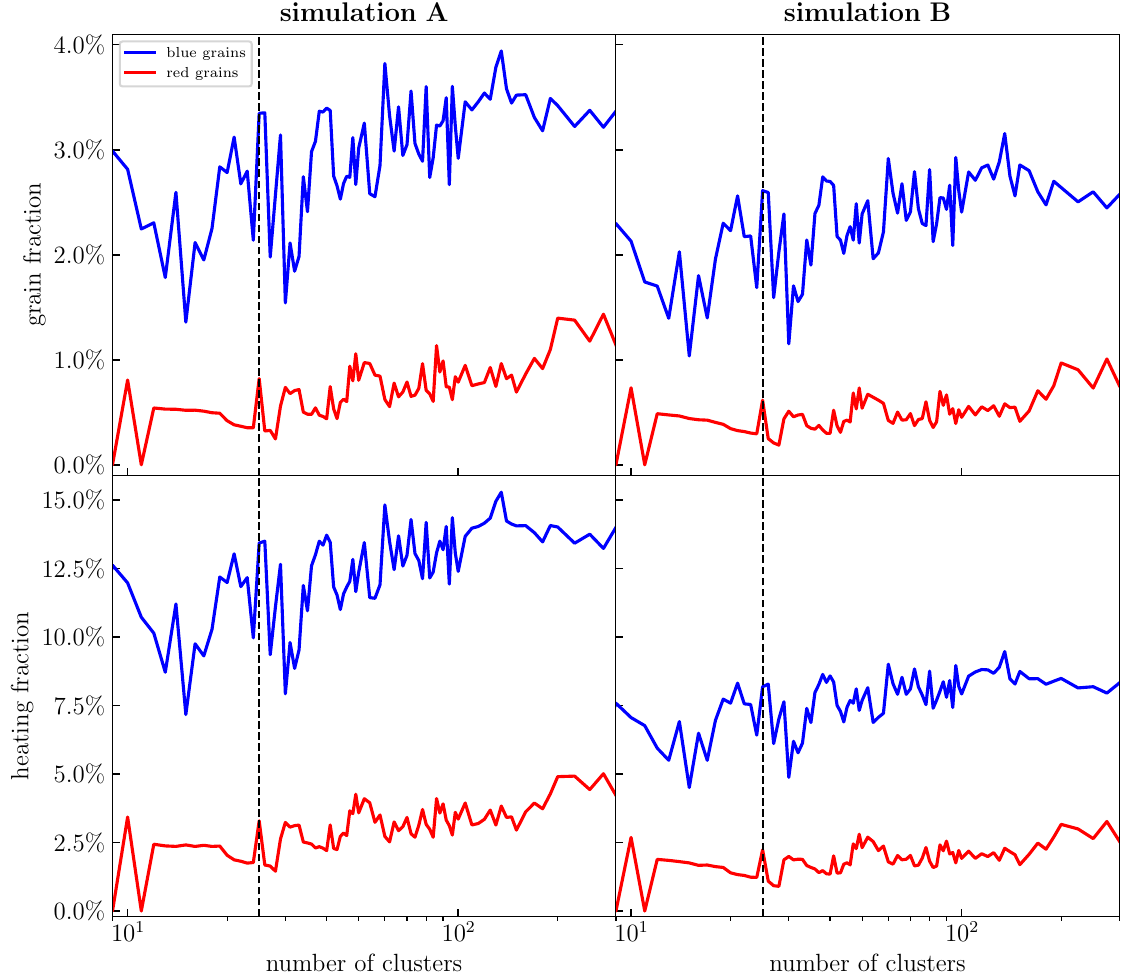}
    \caption{Results from the k-means algorithm versus number of clusters. Top row: percentage of pixels classified as red or blue grains. Bottom row: percentage of the total heating coming from pixels classified as red and blue grains. The vertical dashed lines indicate the number of clusters in the model used in our results.}
    \label{fig:clustering_stats}
\end{figure*}

Figure~\ref{fig:clustering_stats} shows how the total fraction of red and blue grains depends on the number of clusters. The figure shows that the fraction of red blue grains stabilise at a large number of clusters, $n > 10^2$. The dashed line in the figure indicates 25 clusters, which was the model we use in our results. This model gives similar values for blue grain heating as the models using a much higher number of clusters. Using 300 clusters, we obtain a total heating from red and blue grains at 18.2\% for simulation A and 10.9\% for simulation B. This is close to the values we got from using 25 clusters: 16.7\% for simulation A and 10.4\% for simulation B. 

\begin{figure*}
    \sidecaption
    \includegraphics[width=12cm]{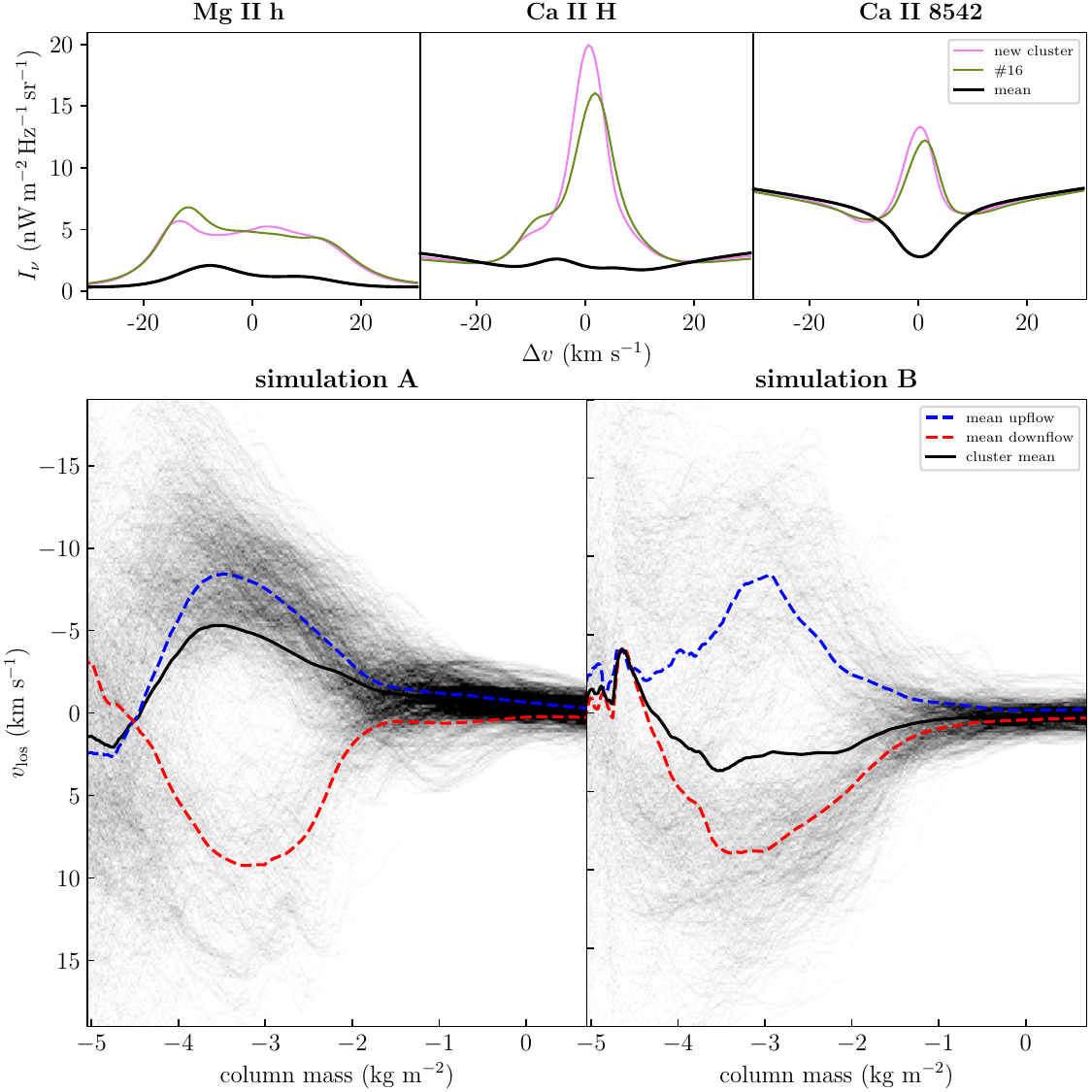}
    \caption{Line-of-sight velocities from a redshifted cluster in emission, from a k-means model with 100 clusters. Top row: representative profile of the new cluster (violet lines) compared to cluster 16 (green lines) in the results. The solid black lines are the mean spectra. Bottom row: Distribution of line-of-sight velocities from each pixel in the new cluster. In the line-of-sight velocity, there are two common distributions: one with down-flows and one with up-flows. The blue and red dashed lines are averages of the up-flows and down-flows, respectively. The solid black line is the average velocity of the cluster.}
    \label{fig:new_velocity}
\end{figure*}

We also investigate the atmospheric stratifications with using a higher number of clusters. The degeneracy in the formation of the spectra in cluster 16 shown in Fig.~\ref{fig:cluster16} also occurred when we clustered the data with 100 clusters. Figure~\ref{fig:new_velocity} shows the line-of-sight velocities from the new cluster that was most similar to cluster 16 in our results. Therefore, the degeneracy in the formation of the slightly red-shifted emission spectra was not an artefact of choosing a small number of clusters.

\end{appendix}

\end{document}